\documentclass{aa}
\usepackage{graphicx}
\usepackage{natbib}
\usepackage{txfonts}
\usepackage{graphicx}
\usepackage{setspace}
\usepackage{xspace}           
\usepackage{sidecap}          
\usepackage{fancyheadings}
\usepackage{lscape}
\def\Msun{M_\odot}
\def\cmtres{\relax \ifmmode {\,\mbox{cm}}^{-3}\else \,\mbox{cm}$^{-3}$\fi}
\def\cmseis{\relax \ifmmode {\,\mbox{cm}}^{-6}\else \,\mbox{cm}$^{-6}$\fi}
\def\ergs{\relax \ifmmode {\,\mbox{erg\,s}}^{-1}\else \,\mbox{erg\,s}$^{-1}$\fi}
\def\kms{\relax \ifmmode {\,\mbox{km\,s}}^{-1}\else \,\mbox{km\,s}$^{-1}$\fi}
\def\ha{\relax \ifmmode {\mbox H}\alpha\else H$\alpha$\fi}
\def\hb{\relax \ifmmode {\mbox H}\beta\else H$\beta$\fi}
\def\hi{\relax \ifmmode {\mbox H\,{\scshape i}}\else H\,{\scshape i}\fi}  
\def\hii{\relax \ifmmode {\mbox H\,{\scshape ii}}\else H\,{\scshape ii}\fi}
\def\oiii{\relax \ifmmode {\mbox O\,{\scshape iii}}\else O\,{\scshape iii}\fi}
\def\oii{\relax \ifmmode {\mbox O\,{\scshape ii}}\else O\,{\scshape ii}\fi}
\def\oi{\relax \ifmmode {\mbox O\,{\scshape i}}\else O\,{\scshape i}\fi}
\def\nii{\relax \ifmmode {\mbox N\,{\scshape ii}}\else N\,{\scshape ii}\fi}
\def\sii{\relax \ifmmode {\mbox S\,{\scshape ii}}\else S\,{\scshape ii}\fi}
\def\lha{\relax \ifmmode \mbox {L}_{H\alpha}\else $\mbox{L}_{H\alpha}$\fi}
\def\ldig{\relax \ifmmode {\mbox L}_{DIG}\else ${\mbox L}_{DIG}$\fi}
\def\ls{\relax \ifmmode {\mbox L}_{ Str}\else ${\mbox L}_{ Str}$\fi}
\def\eme{\relax \ifmmode {\,\mbox{pc\,cm}}^{-6}\else \,pc\,cm$^{-6}$\fi}
\def\l{\relax \ifmmode  \lambda\else $\lambda$\fi}

\def\etal{{et al.~}}
\def\me{$^{-1}$}              
\def\arcmin{\hbox{$^\prime$}}
\def\arcsec{\hbox{$^{\prime\prime}$}}
\def\deg{\hbox{$^\circ$}}
\def\fs{\hbox{$^{\rm s}$}}
\def\hms#1h#2m#3s{\relax \ifmmode #1^{\rm h}\,#2^{\rm m}\,#3^{\rm s}
                   \else \hbox{$#1^{\rm h}\,#2^{\rm m}\,#3^{\rm s}$}
                  \fi}
\def\dms#1d#2m#3s{\relax#1\deg\,#2\arcmin\,#3\arcsec}
\def\hmsd#1h#2m#3.#4s{\relax\ifmmode #1^{\rm h}\,#2^{\rm m}\,#3.#4\fs
                      \else \hbox{$#1^{\rm h}\,#2^{\rm m}\,#3#4\fs$}
                      \fi}
\begin{document}
   \title{The internal dynamical equilibrium of H~II regions: a statistical study}

   \author{M. Rela\~no\inst{1}, J. E. Beckman\inst{1,2}, A. Zurita\inst{3,4}, M. Rozas\inst{5} 
   and C. Giammanco\inst{1}}

   \offprints{M. Rela\~no}

   \institute{Instituto de Astrof\'\i sica de Canarias, C. V\'\i a L\'actea s/n,
       38200, La Laguna, Tenerife, Spain \\
       \email{mpastor@ll.iac.es}, \email{corrado@ll.iac.es}
       \and Consejo Superior de Investigaciones Cient\'\i ficas (CSIC), Spain \\
       \email{jeb@ll.iac.es}
       \and Dpto. de F\'\i sica Te\'orica y del Cosmos, Facultad de Ciencias, U. de Granada,
	Avda. Fuentenueva s/n, 18071, Granada, Spain\\
	\email{azurita@ugr.es}
        \and Isaac Newton Group of Telescopes, Apartado de Correos 321, 38700, Santa Cruz de La Palma, Canarias,
             Spain
	\and Observatorio Astron\'omico Nacional (UNAM), Apartado Postal 877,
             Ensenada, B.C., M\'exico \\
       \email{mrozas@astrosen.unam.mx} 
             }

   \date{}
\authorrunning{Rela\~no et al.}

   \abstract{We present an analysis of the integrated \ha\ emission line 
   profiles for the \hii\ region population of the spiral
galaxies NGC~1530, NGC~6951 and NGC~3359. We show that $\sim$70\% of the 
line profiles show two or three Gaussian components. The relations between 
the luminosity (${\rm log~L_{\scriptsize\ha}}$) and non--thermal line width 
(${\rm log~\rm\sigma_{nt}}$) for the \hii\ regions of the three galaxies are studied and
compared with the relation found taken all the \hii\ regions of the three galaxies as a single distribution. 
In all of these distributions we find a lower envelope in log~$\rm\sigma_{nt}$. 
A clearer envelope in $\rm\sigma_{nt}$ is found when only those \hii\ regions 
with $\rm\sigma_{nt} >\sigma_{s}(13~\kms)$ are considered, where $\rm\sigma_{s}$ is a 
canonical estimate of the sound speed in the interestellar medium. The linear fit for the envelope is 
${\rm log~L_{\scriptsize\ha}}=(36.8\pm0.7)+(2.0\pm0.5)~{\rm log~\rm\sigma_{nt}}$ 
where the \ha\ luminosity of the region is taken directly from a photometric \hii\ region catalogue. 
When the \ha\ luminosity used instead is that fraction of the \hii\ region 
luminosity, corresponding to the principal velocity component, i.e. to
the turbulent non--expanding contribution, the linear fit is
${\rm log~L_{\scriptsize\ha}}=(36.8\pm0.6)+(2.0\pm0.5)~{\rm log~\rm\sigma_{nt}}$, i.e. unchanged but slightly
tighter.
The masses of the \hii\ regions on the envelope using the virial theorem and the mass estimates 
from the \ha\ luminosity are comparable, which offers evidence that the
\hii\ regions on the envelope are virialized systems, while the remaining regions, 
the majority, are not in virial equilibrium.
   }

   \maketitle
%

\section{Introduction}

Supersonic velocity dispersion is a property of the most luminous \hii\
regions which have been extensively studied in the li\-te\-ra\-tu\-re since
Smith \& Weedman (1970) first observed it. In order to find a physical explanation
for the supersonic line widths, Terlevich \& Melnick (1981) proposed that \hii\
regions are gra\-vi\-ta\-tio\-na\-lly bound systems and that the observed velocity
dispersion is produced by motion of discrete ionized gas clouds in the
gravitational field created by the mass distribution inside the \hii\ region.
This conclusion was based on their observational claim that the relations
L~$\sim\sigma^{4}$ and R~$\sim\sigma^{2}$, (where L and R are the luminosity and radius 
of the \hii\ region and $\sigma$ the velocity dispersion of the line profile), 
found in the ste\-llar systems of ellip\-ti\-cal galaxies, bulges
of spiral galaxies and globular clusters, are also found in the gaseous \hii\ regions. 

Several authors have tried to confirm these empirical relations 
but no agreement has been found between the results for the 
$\rm L-\sigma$ relation from the different studies. The variations in the results 
have been attributed to several effects: 1) limitations of the observations 
(Gallagher \& Hunter 1983; Hippelein 1986); 2) comparison between di\-ffe\-rent \hii\ region 
sample criteria (Roy, Arsenault \& Joncas 1986; Arsenault \& Roy 1988); besides, 
the different criteria for the estimates of the radii of the \hii\ regions do not 
allow a valid comparison between the relations found by different authors (Sandage \& Tamman 
1974; Melnick 1977; Gallagher \& Hunter 1983); 3) asymmetries of the integrated \hii\ region line profiles 
(Arsenault, Roy \& Boulesteix 1990; Hippelein 1986), while others have found secondary components 
or asymmetries showing that a single Gaussian fitted to the profile may not be a 
good representation of the inherent velocity dispersion of the gas
(Skillman \& Balick 1984; Rozas \etal 1998), in some cases a Voigt function has better characterized
a significant fraction of the observed line profiles (Arsenault \& Roy 1986).

The most complete study up to now, which 
covers the whole \hii\ region population for a single galaxy, is that by Rozas \etal (1998).
This study did not find the relation $\rm L\propto\sigma^{4}$, but 
obtained a lower envelope in $\sigma$ in the $\rm L-\sigma$ diagram, formed by those \hii\ regions 
which Rozas \etal (1998) suggested are close to virial equilibrium.
Terlevich \& Melnick (1981) found a scatter in the L-$\sigma$ relation which they
suggested might be due to the different metallicities of the \hii\ regions in the sample. 
Hippelein (1986) and Gallagher \& Hunter (1983) suggested that an apparent
dependence of $\sigma$ on metallicity will be induced when using a constant
electron temperature for the co\-rrec\-tion of thermal line broadening, since
T$_{\rm e}$ itself is a function of the metal abundance in the ionized gas. 
The study of abundances in \hii\
regions in the discs of spiral galaxies has shown the existence of 
gradients, with higher abundances towards the centre of a galaxy 
(e.g. Vila-Costas \& Edmunds 1992). These
gradients are small in the discs of barred galaxies, but in ge\-ne\-ral could affect the L--$\sigma$
distribution of the \hii\ regions located across a galaxy disc.

The hypothetical relations of the form L~$\sim\sigma^{4}$ and R~$\sim\sigma^{2}$
obtained for virialized systems are based on the un\-suppor\-ted assumptions for \hii\ regions, that
the ratios M/L and L/R$^{2}$, where M is the mass of the region, are 
constant. Thus, departures from these relations do not provide any evidence
for or against the gravitational equilibrium model (Melnick \etal 1987). 

As shown by Rozas \etal (1998) for the \hii\ regions in NGC~4321, a fiducial test to
prove the virialization of the \hii\ regions must rely on the comparison between
the total mass inside the \hii\ region and the dynamical mass obtained from the
velocity dispersion of the observed line profile using the virial
theorem. 
Such a comparison was made individually for nearby extragalactic \hii\ regions: for NGC~604 in the SMC by
Yang \etal (1996) and for 30~Dor in the LMC by Chu \& Kennicutt (1994). 
Yang \etal (1996) computed the
total mass of NGC~604 and explained using the virial theorem the basic
broadening $\sigma\sim(9-13)$~\kms\ found by them in most positions of the
\hii\ region. Chu \& Kennicutt (1994) could not explain the velocity dispersion of the
central core of 30~Dor as due to virial motions, even though they took into
account not only the io\-ni\-zed mass of the \hii\ region but also an estimate of the neutral 
mass. Rozas \etal (1998) compared the \hii\ region masses obtained from 
the \ha\ luminosity with the masses obtained from the application of 
the virial equilibrium. While the \hii\ regions located well above the 
envelope in $\sigma$ in the L-$\sigma$ diagram, cited above,  
present major differences between the two mass estimates obtained with these 
procedures, the \hii\ regions located on the envelope
show comparable values in the masses derived by the two different methods, 
which was the argument used by Rozas \etal (1998) in claiming that the 
envelope was the locus of virialization.

\vspace{0.5cm}

In this paper we study the L--$\sigma$ relation for the \hii\ regions of three 
nearby Milky Way sized barred spirals, NGC~1530, NGC~3359 and NGC~6951. We know the \ha\ 
luminosity of each \hii\ region from the catalogues obtained from the con\-ti\-nuum 
subtracted \ha\ images of the three galaxies as described in Rozas \etal (1996) 
and Rozas \etal (2000a). The Fabry--P\'erot
observations allow us to extract the line profiles for each \hii\ region in the galaxy
and to obtain the corresponding ve\-lo\-ci\-ty dispersion.

The large number of \hii\ regions in a full
galaxy disc forms a good sample to study the L--$\sigma$ relation because it avoids some
points of the controversy of previous studies. Firstly, all the \hii\
regions are observed with the same instrument, allowing us to extract their
integrated spectra and analyze them with the same procedure. Secondly, since all the \hii\ regions are within a
single galaxy, distance uncertainties in the L--$\sigma$ relation are eli\-mi\-na\-ted for each galaxy. Thirdly, errors
due to using different criteria for estimating the radius of an \hii\
region are reduced because the criteria are the same for all the regions 
and are obtained automatically in the production of the \hii\ region ca\-ta\-lo\-gue for all the objects in the
sample. And finally, we have tried to overcome the difficulties arising from any asymmetries
in the observed line profiles by fitting them with the optimum num\-ber of Gaussian
components, and proposing via specific identification of the minor components, that the 
component which indicates the correct value of 
$\sigma$ is the central most intense component, which in almost all cases contains 
more than 75\% of the total \ha\ luminosity of the region. 

\section{The Observations}

\subsection{Narrow -Band observations}

The data reduction of the continuum subtracted \ha\ images and the production of 
the \hii\ region catalogue for NGC~6951 is reported in
Rozas \etal (1996) and for NGC~3359 in Rozas \etal (2000a). 
Here, we describe briefly the data reduction and extraction of the \hii\ region catalogue
for the third galaxy of our sample, NGC~1530. The parameters of the observations 
for the three galaxies are shown in Table~\ref{tabparamobs}.

\begin{table*}
\hspace{-1cm}
\centering
\caption[]{Observational parameters for narrow--band \ha\ photometric observations
  and the TAURUS Fabry-P\'erot interferometry for NGC~6951, NGC~3359 and
  NGC~1530. Seeing is taken as the FWHM measured in the final continuum-subtracted \ha\
  images of unsaturated stellar sources and the seeing for Fabry-P\'erot observations 
is the FWHM of the continuum images obtained from the analysis of the data 
cubes and the spectral resolution is
  the velocity separation between adjacent planes in the data cube.}
\hspace{-1cm}
\begin{tabular}{l|llll|llll}
\hline
\hline
& \multicolumn{4}{l|}{Narrow--Band} & \multicolumn{4}{l}{Fabry-P\'erot} \\
\hline
& Filter (\ha) & Filter (continuum)  & Pixel Size   & Seeing  & Filter & Pixel Size 
& Seeing & Spect. res. \\
&      &  & (\arcsec/pix)  & (\arcsec) &        & (\arcsec/pix)  & (\arcsec)   & (\kms) \\
\hline
NGC~6951 & 6589/15 & 6565/44  & 0.28 & 0.8    & 6589/15 & 0.58 & 1.3 & 15.63 \\
NGC~3359 & 6594/44 & 6686/44  & 0.59 & 1.5-1.6& 6589/15 & 0.56 & 1.5 & 15.66 \\
NGC~1530 & 6626/44 & HARRIS R & 0.33 & 1.2-1.4& 6613/15 & 0.29 & 1.0 & 18.62 \\
\hline
\end{tabular}
\label{tabparamobs}
\end{table*}

NGC~1530 was observed in \ha\ during the night of August 4th 2001 at the 1m Jacobus 
Kapteyn Telescope (JKT) on La Palma. A Site2 2148x2148 CCD detector was used with a projected pixel 
size of 0.33\arcsec. The observing conditions
were good, with photometric sky and 1.2\arcsec-1.4\arcsec\ seeing. The galaxy was
observed through a 44\AA\ bandwidth interference filter with central wavelength
6626\AA\, quite close to the redshifted \ha\ emission line of the galaxy (6616\AA).
The width of the filter allows the partial transmission of the nitrogen emission
lines [\nii]$\lambda$6548, $\lambda$6583. However, the contribution of these lines to the measured \ha\ 
fluxes accounting for the total intensity of [\nii] lines and the filter transmission at 
the given wavelengths, give rise to a maximum contribution of $\sim$10\% in the measured flux.
For the continuum subtraction, a broad band HARRIS R image was taken. The total
integration times for the on--line and HARRIS R filters were 4800s (4$\times$1200s) and 600s
respectively.

The raw data were processed using standard procedures with the data reduction package IRAF. The images
were bias subtracted and flatfield corrected. Then a constant sky background value was subtracted from
each image of the galaxy, obtained by measuring the sky background in areas of each ima\-ge which were
free of galaxy emission. The on-line and continuum images were then aligned using positions of field
stars in the images and combined separately. Cosmic ray effects
were removed from the final on-line image after combination, via standard rejection operations on each
pixel in the combined images. 

\begin{figure}
\centering
\includegraphics[width=8cm]{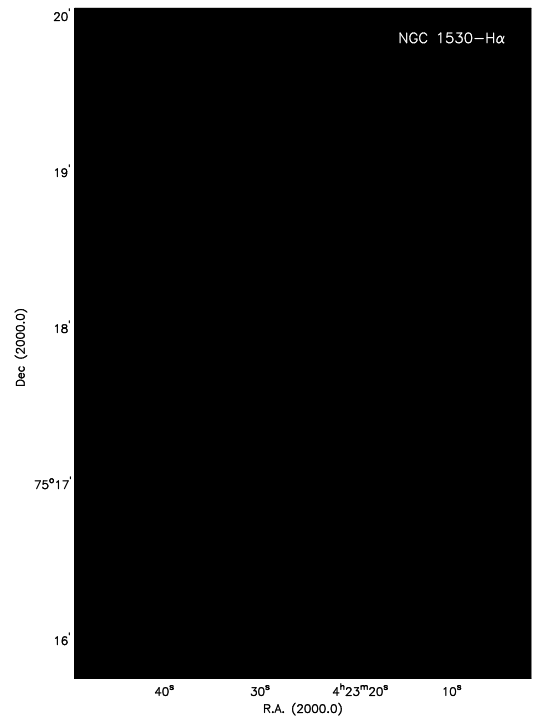}
\protect\caption[ ]{Continuum--subtracted \ha\ image of NGC~1530.}
\label{1530im}
\end{figure}

Finally, the continuum image was subtracted from the on-line image. In order to obtain the scaling
factor for the con\-ti\-nuum subtraction, two different methods were employed. First, we measured the flux
from the field stars in both the on-line and the continuum image. 20 unsaturated stars in the two
ima\-ges yield a mean flux ratio on-band/continuum of 0.020, with a standard deviation of 0.003. Then, we
employed the method described in B\"oker \etal (1999) to get a first estimate of the factor, by dividing 
pixel by pixel the intensity measured in the on-band and continuum images. This method gave a value for
the continuum factor of $0.017\pm0.01$, which overlaps with the previously determined value. We then
generated a set of con\-ti\-nuum subtracted images obtained with different scaling factors ranging from
0.015 to 0.024. A detailed inspection of the ima\-ges showed that the best scaling factor was between 0.017
and 0.019. A factor of 0.018 was finally adopted, with an uncertainty of 5\%. We found in a set of
\hii\ regions that their fluxes
change by less than 4\% due to this uncertainty in the determination of the scaling factor. The
continuum subtracted image of NGC~1530 is illustrated in Fig.~\ref{1530im}. 

Absolute flux calibration of the
galaxy was obtained from the observation of several standard stars from the list of Oke (1990). The
\ha\ luminosity co\-rres\-pon\-ding to one instrumental count was found to be
$5.43\times 10^{35}$\ergs\ ${\rm count}^{-1}$. This factor includes a correction for the fact that the R 
filter employed for the 
continuum subtraction includes the \ha\ emission line of the galaxy. The as\-tro\-me\-try of the reduced image
was performed by identification of foreground stars of the con\-ti\-nuum image in the Palomar Plates. The
accuracy of the as\-tro\-me\-try across the field of view is better than $0.45\arcsec$ (1.4 pixels). 

\subsection{The H\,{\scshape\em II} Region Catalogue}
\label{catalogue}
The production of the \hii\ region catalogue (i.e. the determination of the position, size and luminosity of
the \hii\ regions of a disc of a galaxy) was explained in careful detail 
in Zurita (2001). The catalogue is obtained with the REGION sofware package developed 
by C. Heller and first used in Rozas \etal (1999). The details of the code applied in NGC~3359 can be found 
in Rozas \etal (2000a) and an equivalent though slower procedure for NGC~6951 in Rozas \etal (1996).

The selection criterion for considering an image feature as an \hii\ region is that the feature must 
contain at least an area equal to the spatial resolution of the image in a non-filamentary configuration, 
each one with an intensity of at least three times the r.m.s. noise level of the local 
background above the local background intensity level. 
In Appendix~A we compare this method with methods for defining the extent and the 
lu\-mi\-no\-si\-ty of \hii\ regions 
adopted by some other workers, notably the use of a cut--off at a fixed fraction of the peak brightness. We 
show that the latter yields a fraction of the total luminosity of a region which depends strongly on its 
integrated luminosity, because the brightness gradient varies from centre to edge and varies 
differently from region to region. The variation implied by using a criterion based on the 
noise level gives significantly less variation, even though the noise level will vary between 
observations and observers. We must also bring in as additional evidence a point of consistency 
which supports the use of our 3 times r.m.s noise--cut-off criterion. In a previous 
publication (Zurita, Rozas \& Beckman 2001) we showed that for images deep enough to quantify 
the \ha\ from the diffuse emission in a galactic disc the 3 times r.m.s cut--off above the background level  
used to separate the \hii\ regions from the diffuse component gave excellent and consistent agreement with 
an independent criterion based only on a limiting surface brightness gradient. Thus we 
suggest that a 3 times r.m.s. limit should offer a more reliable way of comparing results among 
different authors than methods based on a fraction of peak surface brightness.

REGION allows us to define as many background regions over the image as necessary. 
The local background for a given \hii\ region is taken as the value of the nearest defined background. 
The r.m.s. noise level of the continuum--subtracted \ha\ image and the adopted selection criteria
establish the lower luminosity limit for detection of the \hii\ regions. 
In the case of NGC~1530 this limit is log~L$_{\scriptsize\ha}=37.80$~(\ergs) for the \ha\ image 
shown in Fig.~\ref{1530im}. The software computes intensity contours from a minimum value and 
after several iterations critical contours mark out the areas in which a set of contiguous 
pixels have intensities higher than three times the r.m.s. noise level above the defined 
background level. The integrated flux, central position and area in pixels of each \hii\ 
region are stored in a file; the radius of the \hii\ region is also given and it is obtained 
assuming that the projected area of the region is the area of a projected sphere on the sky plane.  

\begin{figure}
\centering
\includegraphics[width=8cm]{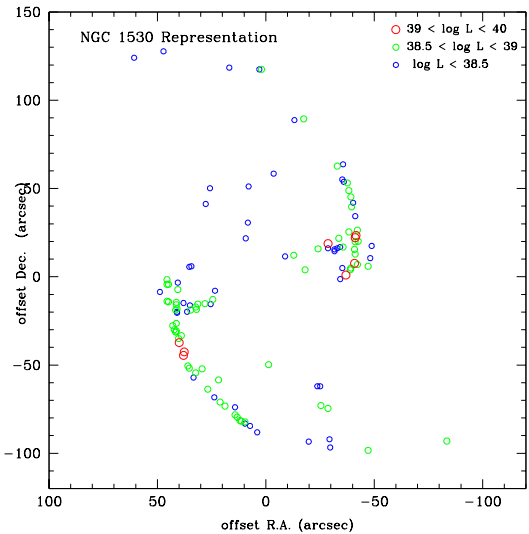}
\protect\caption[ ]{Representation of the position and luminosities of the catalogued \hii\ regions of 
NGC~1530. The coordinates of the centre of the image are R.A.=$\hms 4h23m26.7s$ and DEC=\dms 75d17m44s
(J2000). The ionized regions around the nucleus of the galaxy have not been included in the catalogue, as 
their mutual proximity makes it difficult to delineate them accurately.}
\label{1530cat}
\end{figure}

For NGC~1530, a total of 119 \hii\ regions were catalogued, excluding the nuclear part of the galaxy. A 
representation
of the positions and luminosities of the catalogued \hii\ regions can be seen in Fig.~\ref{1530cat}.

\subsection{Fabry--P\'erot observations}
\label{chap2sec2}
While the photometry from the narrow--band observations gives us a measurement
of the \ha\ luminosity for each \hii\ region in the galaxy and an estimate
of its area, kinematic information can be obtained with
Fabry-P\'erot interferometry. The three barred galaxies, NGC~3359, NGC~6951 and NGC~1530 
were observed 
with the TAURUS-II Fabry-P\'erot interferometer at the 4.2m William Herschel 
Telescope (WHT) on La Palma. The data reduction for each
galaxy is explained respectively in Rozas \etal (2000b) for NGC~3359, in Rozas \etal (2002)
for NGC~6951 and in Zurita \etal (2004) for NGC~1530, and the observational parameters 
are shown in Table~\ref{tabparamobs}. 

The observations consist of exposures at a set of equally stepped separations of the 
etalon, allowing us to scan the full wavelength range of the \ha\ emission line in the galaxy.
For the three galaxies, the observed emission wavelengths were scanned in 
55 steps with exposures times at each position of the etalon of 140s for NGC~3359 
and NGC~6951, and 150s for NGC~1530. An appropriate redshifted narrow--band \ha\ filter 
is used as an order--sorting filter. 
Wavelength and phase calibration were performed using observations of
a calibration lamp before and after the exposure of the data cube. 
The final ca\-li\-bra\-ted data cube has three axes, $x$ and $y$, corresponding to the 
spatial coordinates in the field, and $z$, which corresponds to the wavelength 
direction. The free spectral ranges for each galaxy are 17.22\AA\ 
for NGC~3359 and NGC~6951 and 18.03\AA\ for NGC~1530, with a 
wavelength interval between consecutive planes of 0.34\AA\ for 
NGC~3359 and NGC~6951 and 0.41\AA\ for NGC~1530. These values give a 
finesse of 21.5 for NGC~3359 and NGC~6951 and 21.2 for NGC~1530, so 
that our 55 planes imply a very slight oversampling. We checked 
that there is no contamination of the atmospheric OH lines in our observations. 

The continuum of the observational data cube was determined by fitting 
a linear relation to the line--free channels, and was then subtracted from the 
line emission channels. From the continuum subtracted data cube we can obtain a detailed 
des\-crip\-tion of the spatial distribution of intensities and velocities provided by 
the moment maps: the total intensity (zeroth moment), velocity (first moment) and 
velocity dispersion (second moment) maps. The general procedure for obtaining the moment 
maps is explained in Knapen (1997). The intensity map is used to identify 
each \hii\ region in the catalogue with the equivalent \hii\ region in the data cube, as we explain in the
next section. 
 
\section{Relating the photometric and spectroscopic data sets}

In order to assign the \ha\ luminosity of each \hii\ region to its 
corresponding line profile, we need to identify each \hii\ region in the 
Fabry-P\'erot data cube with its corresponding region in the
continuum subtracted \ha\ flux calibration image (Fig.~\ref{1530im} 
for NGC~1530). There are some problems involving the identification of the \hii\ regions: 
\begin{itemize}
\item[1.-] The pixel scales of the CCDs used in the photometric 
observations and those used in the observations with the Fabry--P\'erot interferometer 
are not the same (see Table~\ref{tabparamobs}). Thus, a region in the {\it calibration 
image}\footnote{We refer to the continuum subtracted \ha\ image from which the 
\hii\ region catalogue has been obtained as the {\it calibration image}. The integrated 
map (zeroth moment map) obtained as a sum of the profile amplitudes in each pixel of the continuum
subtracted \ha\ data cube is here called the {\it intensity map}.} will be
scaled differently from the intensity map extracted from the data cube. 
\item[2.-] Not all of the \hii\ regions in the catalogue derived from the calibration image
appear in the intensity map, which is used to select the 
aperture for the extraction of the line profiles in the continuum subtracted data cube. This is because 
the total exposure
time for each plane in the data cube is shorter than the exposure time for
the \ha\ image and the Fabry--P\'erot throughput is less than that of the direct
ima\-ging cameras.
\end{itemize}

With these considerations in mind the identification of each
\hii\ region was made interactively on the computer screen. Once we
recognized a given 
catalogued \hii\ region in the intensity map we fitted a two-dimensional Gaussian function to 
the emission feature and obtained the centre for the \hii\ region in the intensity map. With this 
procedure we obtain a file containing, for each \hii\ region detected in the Fabry--P\'erot observations,
its coordinates in pixels, and its luminosity and radius assigned in the
catalogue. In this way, we assign an integrated \ha\ flux to the \hii\ region spectrum obtained from
the data cube and relate the photometric data with the kinematic information
extracted from the interferometric observations. The following notes for each
galaxy clarify the procedure for the identification of the \hii\ regions:

\begin{itemize}

\item{\bf NGC~1530}: As shown in Table~\ref{tabparamobs}, the pixel scales for the 
calibration image and the intensity map are similar, so that most of the
\hii\ regions are quite easily
identified in both the calibration image and the intensity map of the Fabry--P\'erot. However, 
there are some catalogued \hii\ regions which appear with 
only a faint signal in the data cube. These \hii\ regions are too faint to be
detected as an \hii\ region feature in the intensity map, only those which
are isolated could be identified. Although we could not fit a centre to 
these features in the intensity map, we could extract the line profile of the faint emission zone 
because the continuum subtracted data cube  
contained sufficient signal. In these cases a valid \hii\ region spectrum could be extracted 
and analyzed, but in the remaining faint regions this was not possible. Besides, 
the projected CCD size for the intensity map 
was too small to cover the whole of the spiral arms, so that some \hii\ regions
at the ends of both spiral arms which are catalogued in the calibrated image 
do not appear in the intensity map and could not be identified. 

There are a few cases in which a single emission zone in the intensity map
corresponds to more than one \hii\ region in the catalogue. We have found 6
faint emission zones in which this situation appears. For these cases the
spectrum of the central component was assigned to the most luminous \hii\ 
region inside the emission zone. Of the 119 \hii\ regions which form the 
catalogue of NGC~1530, 98 could be reliaby identified in the Fabry-P\'erot data, 
the remainder belonging to one of the categories described above.

\item{\bf NGC~3359}: Apart from 20 \hii\ regions that are outside the field of the
intensity map the rest of the 202 \hii\ regions down to a value of log~L$_{\scriptsize\ha}=37.7$~(\ergs)
could be easily identified in the data cube. This exact lower limit of identification is somewhat
arbitrary. Although the lower limits to the luminosity of 
the detected \hii\ regions in the calibration image is log~L$_{\scriptsize\ha}=36.70$ 
(\ergs), the line profiles extracted for \hii\ regions with an \ha\ 
luminosity below log~L$_{\scriptsize\ha}=37.7$~(\ergs)
have velocity dispersions with large errors (see
Table~\ref{errLsigma}). Besides, the present study is confined to
the most luminous \hii\ regions for each galaxy, so we consider
that the information from the faintest \hii\ regions in the study of the
log~L$_{\scriptsize\ha}$--log~$\rm\sigma_{nt}$ diagrams is of limited value and importance in this
context.

\item{\bf NGC~6951}: Down to a limit of log~L$_{\scriptsize\ha}=38.0$~(\ergs), there are 91 \hii\ regions 
ca\-ta\-lo\-gued and we could identify 73 in the intensity map. We have found 7
cases in which more than one \hii\ region corresponds to a
single emission zone in the intensity map. This is due to the large diffe\-ren\-ce in
pixel scales between the intensity map and the \ha\ calibration image (0.28\arcsec/pix, for the 
photometric data and 0.58\arcsec/pix for the Fabry--P\'erot observations, see Table~\ref{tabparamobs}). 
We considered in detail only 4 of these cases (regions 4, 5, 17, and 26 in the
catalogue); the other 3 were discarded because they are small, faint
regions having more than two catalogued \hii\ regions inside the emission feature. 
In the 4 \hii\ regions 4, 5, 17, and 26 more intense knots could be
identified inside the emission zones related to the \hii\
regions catalogued and the line profiles were extracted for these knots.

Taking into account the problems involved in the identification of the \hii\ 
regions in the three galaxies described above, we finally used data from 98 
regions in NGC~1530, 202 in NGC~3359 and 73 in NGC~6951. 

\end{itemize}

\section{Extraction of the line profiles}

Once we have identified the \hii\ regions for each galaxy in each
corresponding intensity map, we obtained the
integrated spectrum of each one from the continuum subtracted \ha\ data
cube. The selection of the aperture that defines the integrated spectrum for 
the \hii\ regions requires special attention. In the case of an isolated
\hii\ region the radius defined in the catalogue gives us an idea of the
radius for the best aperture. But in the case of a non-isolated \hii\ region
or a region that is not spherical (especially those located
along the bar), the radius is taken here as a reference value to identify 
and separate it from its neighbours.

The error in the velocity dispersion of the most intense, central, component when different
apertures are chosen to extract the line profile, is shown in Table~\ref{taberrsigma}. In this
table we have selected as conservative examples, three non-isolated \hii\ regions from each galaxy of 
our sample. We extracted their line profiles for different apertures: two
rectangular apertures with different sizes, one bigger than the other,
and a circular aperture. The
apertures are specified in Table~\ref{taberrsigma}, the radius for the
circular aperture is the mean value of the radii obtained from the rec\-tan\-gu\-lar
apertures, assuming their areas are equal to those of circular projected areas.

As shown in this table, the velocity dispersion of the central, most intense,
component obtained using different apertures does not change significantly
and the changes are within the error bars of $\sigma$. The errors in velocity
dispersion shown here are those given by the fit program when a least
squares fit is applied to the observed function using the proposed function
formed by Gaussian components. For all the \hii\ regions listed
in Table~\ref{taberrsigma}, the variation in $\sigma$ obtained with the
different selected apertures is comparable with the estimate given by the
fitting program. In two cases (regions 11 and 35) these variations are
significantly larger than the rest. The cause is somewhat different for
the two cases. For region 11 there are two strong peaks (one significantly
stronger than the other), which leads to a larger error in $\sigma$;
for region 35, which is the faintest chosen region in NGC~1530, the error
is larger due to an inferior S:N ratio. The examples shown in
Table~\ref{taberrsigma} give us some freedom to take the aperture that we think
can best specify the \hii\ region, which can be hard to judge when the \hii\
regions are crowded. For simplicity, we chose rectangular apertures and
extracted the integrated profiles with the task {\sc profil} in the GIPSY
package.

The selected regions taken as examples in Table~\ref{taberrsigma} 
are not isolated but embeded in crowded fields of \hii\ regions, so the
results in Table~\ref{taberrsigma} give estimates of the upper limit to 
the uncertainty in the velocity dispersion for the central component when
using different apertures. For  
isolated \hii\ regions the uncertainties will be lower because there is no
contamination by emission from neighbouring regions. In these cases the 
aperture is taken to include all the  
emission defined in the intensity map.  

\begin{table*}
\centering
\caption[]{Estimates of the uncertainties in the velocity dispersion due to
  changes in aperture dimensions for the
  central and most intense Gaussian  
function fitted to the observed spectrum for a group of selected \hii\
regions in the barred galaxies: NGC~3359, NGC~1530 and NGC~6951. The catalogued radii of the \hii\ regions (Rad. in column~3) 
and the apertures are in pixels in the intensity maps and $\sigma$ in
\kms. The third value for $\sigma$ is obtained with a circular aperture of radius r. This radius 
r is the mean value of the radii obtained from the rectangular
apertures, assuming their areas are equal to those of circular projected areas.}
\begin{tabular}{ccc|cc|cc|cc}
\hline
\hline
Region & log~L$_{\scriptsize\ha}$ & Rad. & $\sigma$ & Aperture & $\sigma$ & Aperture & $\sigma$ & r \\
(galaxy) & (\ergs) & (pix) & (\kms) & (pix$\times$pix) & (\kms) & (pix$\times$pix) & (\kms) & (pix) \\
\hline
11(N1530)  & 39.4338 & 6.0 &  39.5$\pm$5.0 &15$\times$8 & 38.6$\pm$4.0 & 7$\times$5 &
41.5$\pm$4.3 & 4.8 \\
15(N1530)  & 39.3467 & 5.9 &  40.7$\pm$0.7 & 7$\times$7 & 41.0$\pm$0.6 & 6$\times$4 &
40.4$\pm$0.6 & 3.4 \\
35(N1530)  & 38.8842 & 3.8 &  35.3$\pm$5.1 & 7$\times$8 & 34.6$\pm$3.3 & 6$\times$4 &
38.8$\pm$3.8 & 3.5 \\
63(N3359)  & 38.4188 & 5.9 &  26.0$\pm$0.9 & 7$\times$5 & 25.2$\pm$0.8 & 4$\times$5 &
25.4$\pm$0.8 & 3.3 \\
51(N3359)  & 38.5170 & 5.1 &  27.1$\pm$0.8 & 8$\times$6 & 25.6$\pm$0.8 & 6$\times$5 &
26.7$\pm$0.9 & 3.5 \\
41(N3359)  & 38.5965 & 6.1 &  28.4$\pm$0.5 & 8$\times$11 & 28.3$\pm$0.5 & 7$\times$6 &
28.7$\pm$0.5 & 4.5 \\
21(N6951)  & 38.5823 & 2.9 &  31.0$\pm$0.8 & 7$\times$8 & 31.1$\pm$0.7 & 5$\times$5 &
31.2$\pm$0.6 & 3.5 \\
28(N6951)  & 38.4441 & 2.6 &  24.8$\pm$0.8 & 8$\times$6 & 24.7$\pm$0.7 & 4$\times$5 &
24.8$\pm$0.6 & 3.2 \\
3(N6951)   & 39.1203 & 4.4 &  29.1$\pm$0.5 & 7$\times$8 & 28.7$\pm$0.5 & 6$\times$5 &
28.9$\pm$0.4 & 3.6 \\
\hline
\end{tabular}
\label{taberrsigma}
\end{table*}

\section{Spectral analysis}

\begin{figure*}
\centering
\vspace{2cm}
\includegraphics[width=15cm]{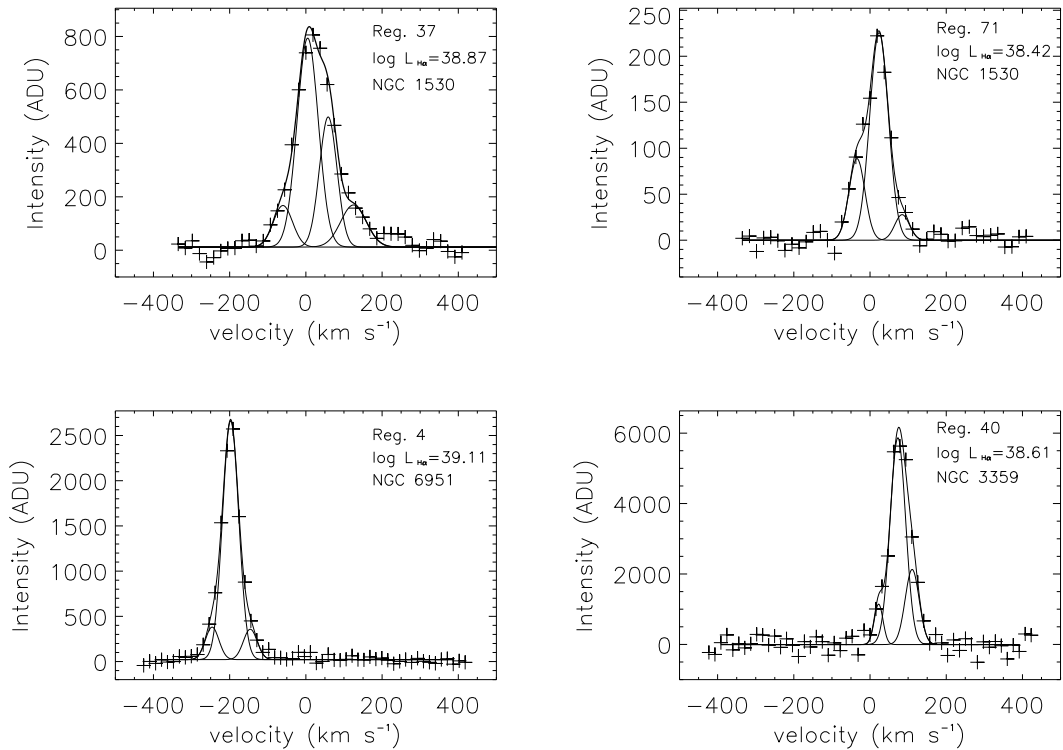}
\caption[]{Integrated line profiles for four \hii\ regions of the three galaxies: NGC~1530, NGC~3359 and
NGC~6951. The spectral resolutions in the line profiles are 18.6~\kms\ for NGC~1530, and 15.6~\kms\ for
  NGC~3359 and NGC~6951.}
\label{especintset}
\end{figure*}

For each \hii\ region, the line profile was extracted taking an integration 
of the selected zone in each plane of the con\-ti\-nuum subtracted data cube.  
The number of Gaussians for each decomposition was given by the requirements 
of the fit program in each line profile. The fits were performed with the task {\sc profit} in the 
GIPSY program, which lets us to fit for each spectrum up to five 
Gaussian components. In Fig.~\ref{especintset} we show some  
integrated line profiles fitted with the optimized number of Gaussians.

Although, the decomposition of the spectra into Gaussian functions is quite clear for
most of the line profiles, there is a minority of cases in which the number of
Gaussians required to fit the spectrum can be ambiguous. We treated these cases carefully for each galaxy.

In NGC~3359 there are 21 spectra (11.5\% of the total number of identified \hii\ regions) that 
have single isolated wavelength points with high intensity values located near the central 
component. To check the possible effect of these on the line widths of the central components, we 
fitted these spectra with a single and with two Gaussian components. In just 6 cases the values
of the velocity dispersion of the principal component for the line profiles
fitted with one or with more
Gaussians do not coincide within the error bars, but the differences are really
small\footnote{The mean of the errors in the velocity dispersions of the central
components for the two type of fits represents $\sim$40\% of the difference
between the velocity dispersion values.}. In addition, NGC~3359 has 7
(3.8\%) regions whose line profiles show low intensity components which are not clearly defined.
Except for 2 cases the velocity dispersion
of the single Gaussian fitted to the spectrum coincides within the
error bars with the values of the velocity dispersion of the central peak
when the spectrum is fitted with more than one Gaussian, and the differences
even in these 2 cases are negligible. In NGC~6951 there are 6 (8.2\%) ambiguous 
fits and in all of them the velocity dispersions associated with the central peak coincide for the
two possible fits within the error bars. In NGC~1530 we do not find ambiguous line 
profiles, because of the high S:N ratio of its line profiles. 

These apparent ambiguities give an idea of the quality of the fit; 
those line profiles which do not have clear multiple Gaussian components give values
of the velocity dispersion for the central peak which do not differ
significantly when a single or multiple Gaussian fit is applied, because the subsidiary 
peaks are fairly weak, and well shifted in wavelength from the line centre. 

\subsection{Statistics of the number of Gaussian components fitted to each
  spectrum}

In Table~\ref{tabcomp} we list, for each galaxy, the fraction of the line profiles with different
Gaussian decompositions. Most of the line profiles ($\sim$70\%)
have two or three Gaussian components. This result is similar to that found by
Arsenault \& Roy (1996) for 47 giant extragalactic \hii\ regions in 26 nearby
galaxies. They found that 43\% of the spectra had profiles 
characterized by a Voigt profile and 21\% were fitted with two Gaussians. The Voigt profile
would correspond better in our case to three Gaussian components, a central peak and
two symmetric low intensity Gaussians. 
   
\begin{table}
\centering
\caption[]{Spectral Gaussian decomposition frequency analysis for the integrated line
        profiles extracted for the identified \hii\ regions in NGC~1530, NGC~3359
        and NGC~6951. Each column shows the fraction of the total emission
        profiles which are fitted by the specified number of Gaussian components (GC).}
\begin{tabular}{cccccc}
\hline
\hline
    Galaxies & 1~GC & 2~GC & 3~GC & 4~GC & 5~GC      \\
             \hline
    NGC~1530 &    10.2\% & 21.4\% & 50.0\% &18.4\% &1.0\%   \\
    NGC~3359 &    26.9\% & 47.3\% & 24.7\% & 1.0\% & --      \\
    NGC~6951 &    25.0\% & 45.0\% & 26.0\% & 4.0\% & --      \\
\hline
\end{tabular}
\label{tabcomp}
\end{table}

Although we will not attempt in this paper to find a physical explanation for all the secondary components
that appear in the line profiles, it is interesting to note that most of the line profiles are characterized by a
central peak and one or two high velocity features, and just a small fraction
contain two central peaks of comparable amplitude.  In order to show this statement we have 
classified the integrated spectra for all the identified \hii\ regions in a qualitative 
form. The results are shown in Table~\ref{tabspecclas}. When 
there is a single Gaussian, this component describes the peak of the line profile. Secondary Gaussian 
components located at more than $\sim$~35~\kms\ 
from the centre of the more intense Gaussian are taken to be {\it
  wing}\footnote{The wings are the outer regions of spectral lines. Here we
  call {\it wing features} the subsidiary spectral emission lines having 
  lower intensity than the principal component which are observed at relatively high velocity to the red
  and/or to the blue of the main emission peak.} features. For example, a spectrum can have two Gaussians
describing the central peak (2~P) or three components, one describing the
central peak and the others classified
as wing features (1~P+2~W). It is interesting to note that there is a relation 
between the fraction of the multiple component Gaussians in the 
central peak and the inclination of the galaxy. The inclinations of the 
galaxies are 55\deg, 53\deg and 42\deg, for NGC~1530, NGC~3359 
and NGC~6951. The most inclined galaxy has the biggest fraction of 
line profiles classified as 2~P. This means that there is a crowding 
effect in the \hii\ region line profiles which grows principally with increasing inclination.

\begin{table*}
\centering
\caption[]{Frequency distribution of the component decomposition for the integrated line
        profiles extracted for the selected \hii\ regions in NGC~1530, NGC~3359
        and NGC~6951. P; central peak or component included in the central
        peak, W; high velocity low intensity {\it wing} feature.}
\begin{tabular}{cccccccc}
\hline
\hline
             & 1~P& 1~P+1~W & 1~P+2~W & 2~P & 2~P+1~W & 2~P+2~W & 2~P+3~W       \\
             \hline
    NGC~1530 &  9.2\% &  8.2\% & 30.5\% & 14.3\% &18.4\% & 18.4\% & 1.0\%   \\
    NGC~3359 & 28.0\% & 35.7\% & 18.1\% & 12.1\% & 5.0\% &  1.1\% & --      \\
    NGC~6951 & 25.0\% & 41.0\% & 27.0\% & 3.0\%  & --    &  4.0\% & --     \\
\hline
\end{tabular}
\label{tabspecclas}
\end{table*}

\section{The luminosity v. non--thermal velocity dispersion relation}

\subsection{Construction of the diagrams}
\label{elimsigmas}

From the integrated spectra of the identified \hii\ regions in the three galaxies NGC~3359, NGC~6951 and 
NGC~1530, we have selected the widths of the most intense Gaussian components. 
These observed widths were corrected for the natural, thermal and instrumental line
widths. The non--thermal velocity dispersion was obtained from the following
expression:

\begin{equation}
\sigma_{\rm nt}=((\sigma_{\rm obs})^{2}-(\sigma_{\rm n})^{2}-(\sigma_{\rm
  t})^{2}-(\sigma_{\rm inst})^{2})^{1/2}
\end{equation}
where $\sigma_{\rm obs}$, $\sigma_{\rm n}$, $\sigma_{\rm t}$ and
$\sigma_{\rm inst}$ are the observed, natural, thermal and
instrumental widths, respectively. The natural width has a value of 3~\kms\
(O'Dell \& Townsley 1988). The thermal width was taken assuming a constant temperature 
inside the \hii\ region of $10^4$~K, which
corresponds to 9.1~\kms (Osterbrock 1989). The instrumental width for each galaxy was obtained from the data cube taken
with the calibration lamp. We extracted a line profile from this calibration
data cube centred on the optical centre of the instrument, where the line
profile of best S:N was obtained. The instrumental width is the half-width,
$\sigma$, of the single Gaussian component fitted to the instrumental line
profile. The values we found for each observation are (17.4$\pm$0.4)~\kms,
(16.3$\pm$0.3)~\kms and (15.9$\pm$0.3)~\kms for NGC~1530, NGC~3359 and NGC~6951,
respectively. The instrumental line profile was analyzed, extracting line
profiles with different apertures at different distances from
the optical centre of the instrument; differences in the instrumental widths 
were no larger than ~10\% across the field of view.

In order to construct the log~L$_{\scriptsize\ha}$--log~$\rm\sigma_{nt}$ diagram, we need to find the 
luminosity associated with the central most intense component of the integrated spectrum. In the case where the line profile is fitted by a single 
Gaussian the assignment is straightforward, but when the line profile is fitted with more components 
there is an uncertainty in the assignment of the luminosity. This is because the 
luminosity of the catalogue co\-rres\-ponds to the integrated luminosity of the \hii\ region identified with the REGION
program, while when a line profile is fitted by more than 
a single Gaussian, it is the sum over the areas of all the components 
which represents the integrated luminosity given by the catalogue. 

To cope with this problem, we have treated specially the \hii\ regions with two clearly 
defined central peaks and with the most intense one having less than 80\%
of the total area below the fitted Gaussians. To these \hii\ regions we assigned the fraction of the 
luminosity of the \hii\ region in the catalogue co\-rres\-pon\-ding to the fractional integrated intensity of the 
strongest Gaussian, while to the other regions we assigned the full value of the catalogue luminosity. 

\begin{figure*}
\centering
\includegraphics[width=12cm]{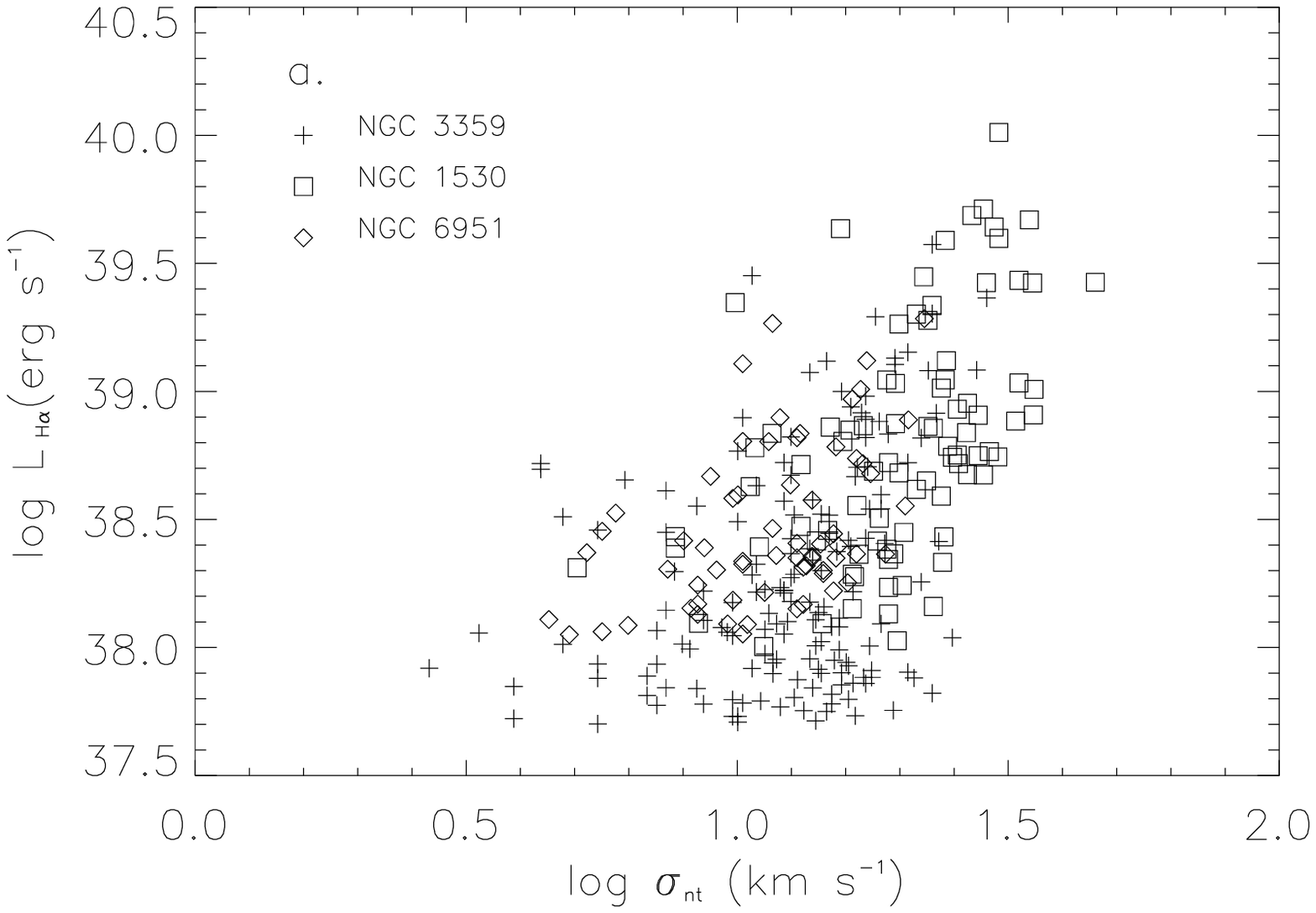}
\includegraphics[width=12cm]{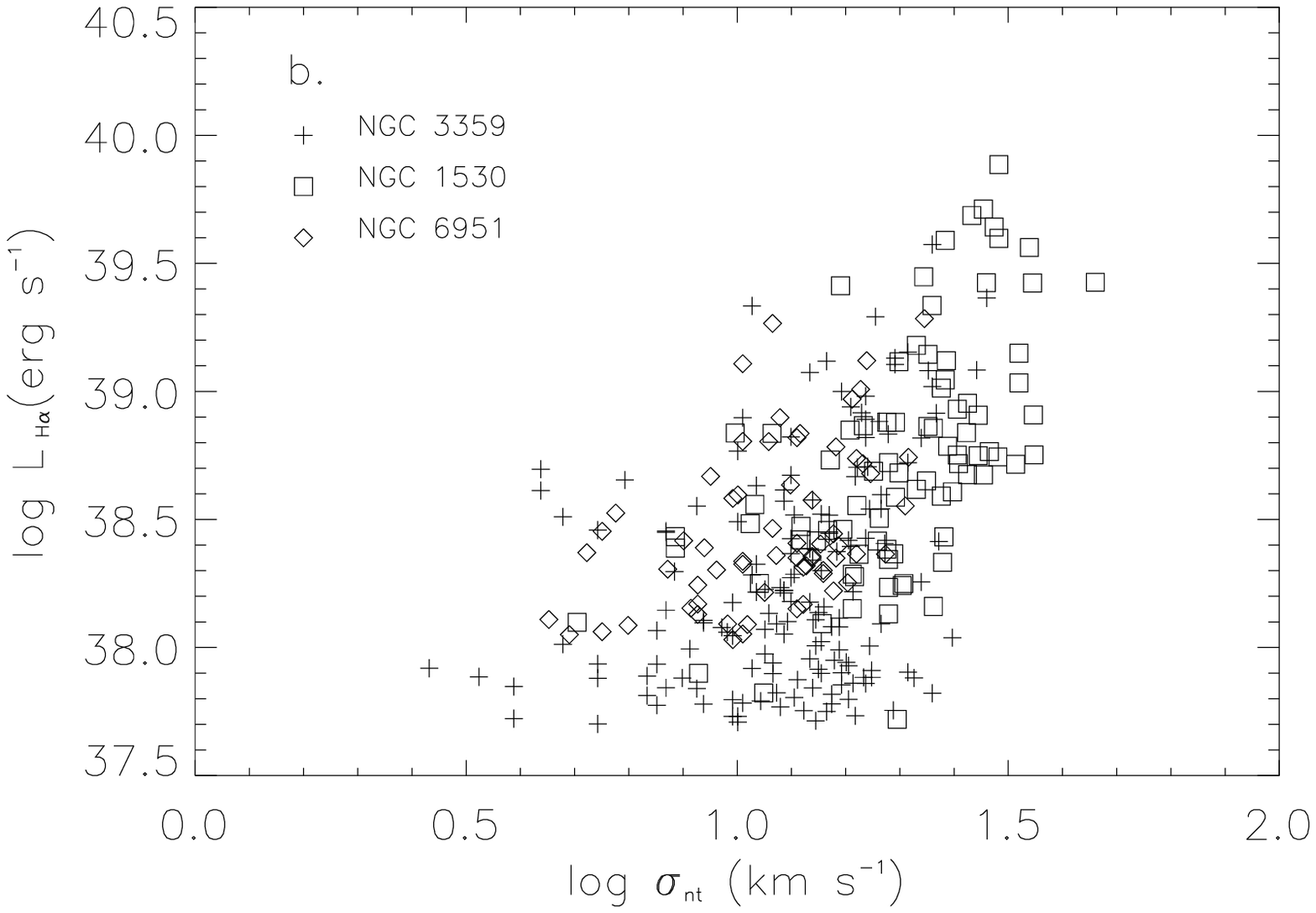}
\protect\caption[ ]{log~L$_{\scriptsize\ha}$--log~$\rm\sigma_{nt}$ diagram for the \hii\ regions in NGC~1530,
    NGC~3359, and NGC~6951. In a. the \ha\ luminosity of the \hii\ region is the luminosity 
    given by the catalogue, while in b. the \ha\ luminosity is the fraction 
    of the \ha\ luminosity from the catalogue corresponding to the fractional
    integrated intensity of the strongest Gaussian component in the \hii\
    region spectrum. $\rm\sigma_{nt}$ is the non--thermal velocity dispersion corresponding to the most 
    intense central component.}
\label{Lsigmatres}
\end{figure*}

In Fig.~\ref{Lsigmatres}a we show the logarithmic \ha\ luminosity of the \hii\
regions in the three galaxies {\it versus} the logarithmic non--thermal velocity dispersion
of the central most intense component of their integrated spectra. In
Fig.~\ref{Lsigmatres}b we show the same 
distribution but, where appropiate, using the fractional \ha\ luminosity corresponding to the central peak, as explained 
above. 

The diagrams show that the range of non--thermal velocity dispersions at a given
lu\-mi\-no\-si\-ty decreases with increasing luminosity of the \hii\ region, i.e. the
diagram becomes narrower for higher luminosities. In both
Figs.~\ref{Lsigmatres}a and~\ref{Lsigmatres}b
it is easy to see that there is an observed upper limit in log~L$_{\scriptsize \ha}$, i.e. 
for a given $\rm\sigma_{nt}$; there is an observed upper limit in 
log~L$_{\scriptsize \ha}$  where no \hii\ regions with lower velocity dispersion are found.

It is interesting to note the ranges of uncertainty in the non--thermal velocity dispersion
of the \hii\ regions in these dia\-grams. The uncertainties are the
error bars in $\sigma_{\rm nt}$ given by the fit procedure and are related to the
signal to noise ratio of the line profiles. We have divided the diagrams into 
two zones, separated by a value of the non--thermal velocity dispersion of
13~\kms, a canonical estimate of the sound speed in the interestellar medium. The relative errors in velocity dispersion for \hii\ regions with
$\rm\sigma_{nt}<13$~\kms are $\sim$28\%, while for \hii\ regions with
$\rm\sigma_{nt}>13$~\kms the relative errors are $\sim$4\%, as shown in
Table~\ref{errLsigma}. The difference in the mean relative errors 
is because for $\rm\sigma_{nt}<13$~\kms, the observed line width is 
appro\-aching the resolution of the instrument, and the S:N is also 
falling because the regions have lower luminosity.

\begin{table*}
\centering
\caption[]{Fraction (in \%) of the number of \hii\ regions for each galaxy with
  $\rm\sigma_{nt}<13$~\kms and $\rm\sigma_{nt}>13$~\kms and the mean value of the relative errors in the velocity dispersion
of the central components for these two groups of \hii\ regions.}
\begin{tabular}{ccccc}
\hline
\hline
& \multicolumn {2}{c}{\hii\ regions with $\rm\sigma_{nt}>13$~\kms} 
& \multicolumn {2}{c}{\hii\ regions with $\rm\sigma_{nt}<13$~\kms} \\
    Galaxy   & Fraction & $\Delta\rm\sigma_{nt}$ & Fraction & $\Delta\rm\sigma_{nt}$ \\
\hline 
    NGC~1530 & 88.2\% & 4.1\% & 11.8\% & 32.1\% \\
    NGC~3359 & 54.4\% & 4.6\% & 45.6\% & 24.9\% \\
    NGC~6951 & 48.4\% & 3.8\% & 51.6\% & 27.8\% \\
\hline
\end{tabular}
\label{errLsigma}
\end{table*}

\subsection{Envelope of the log~L$_{\scriptsize\ha}$--log~$\rm\sigma_{nt}$ diagram}
\label{envelope}
\begin{figure*}
\centering
\includegraphics[width=12cm]{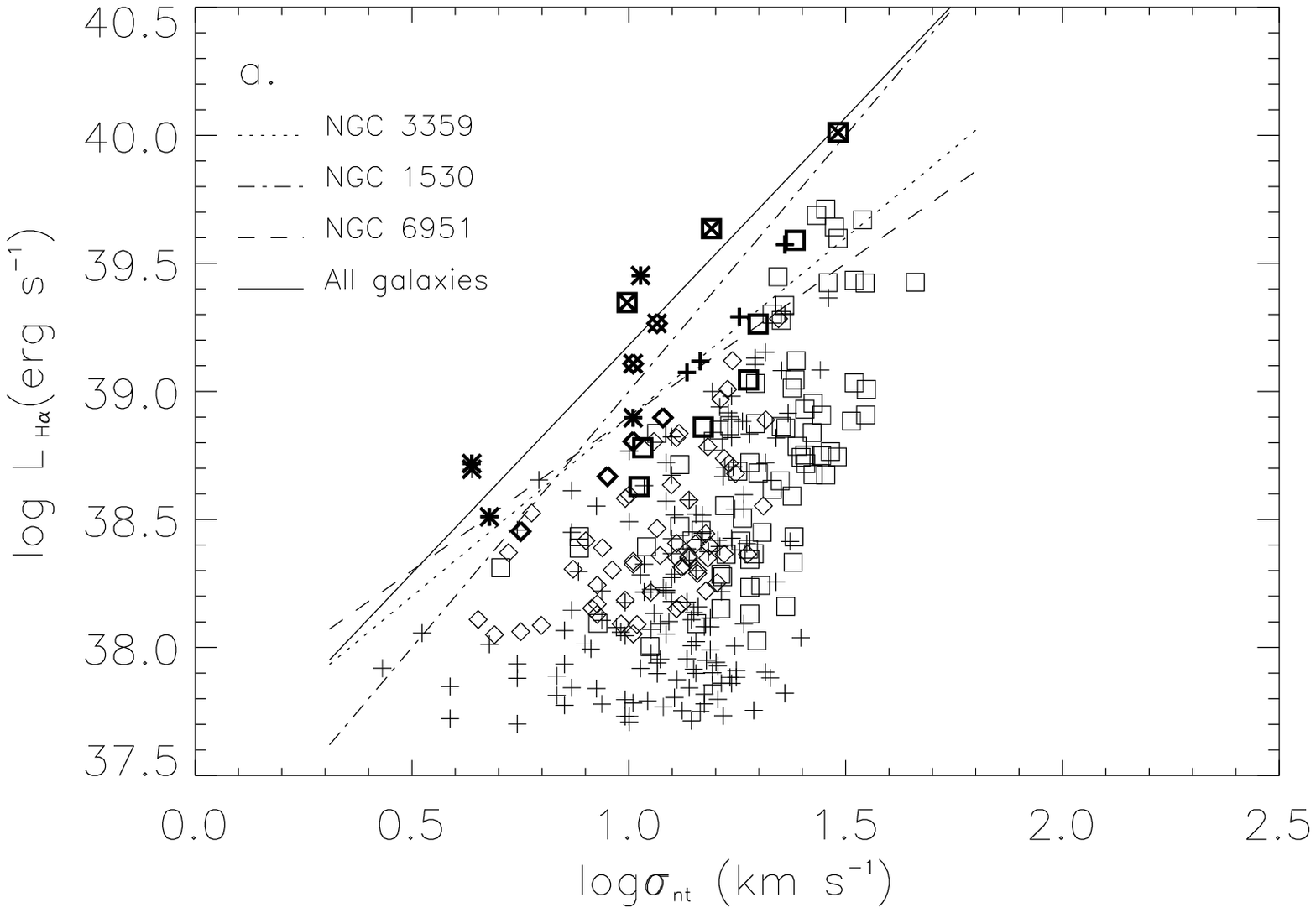}
\includegraphics[width=12cm]{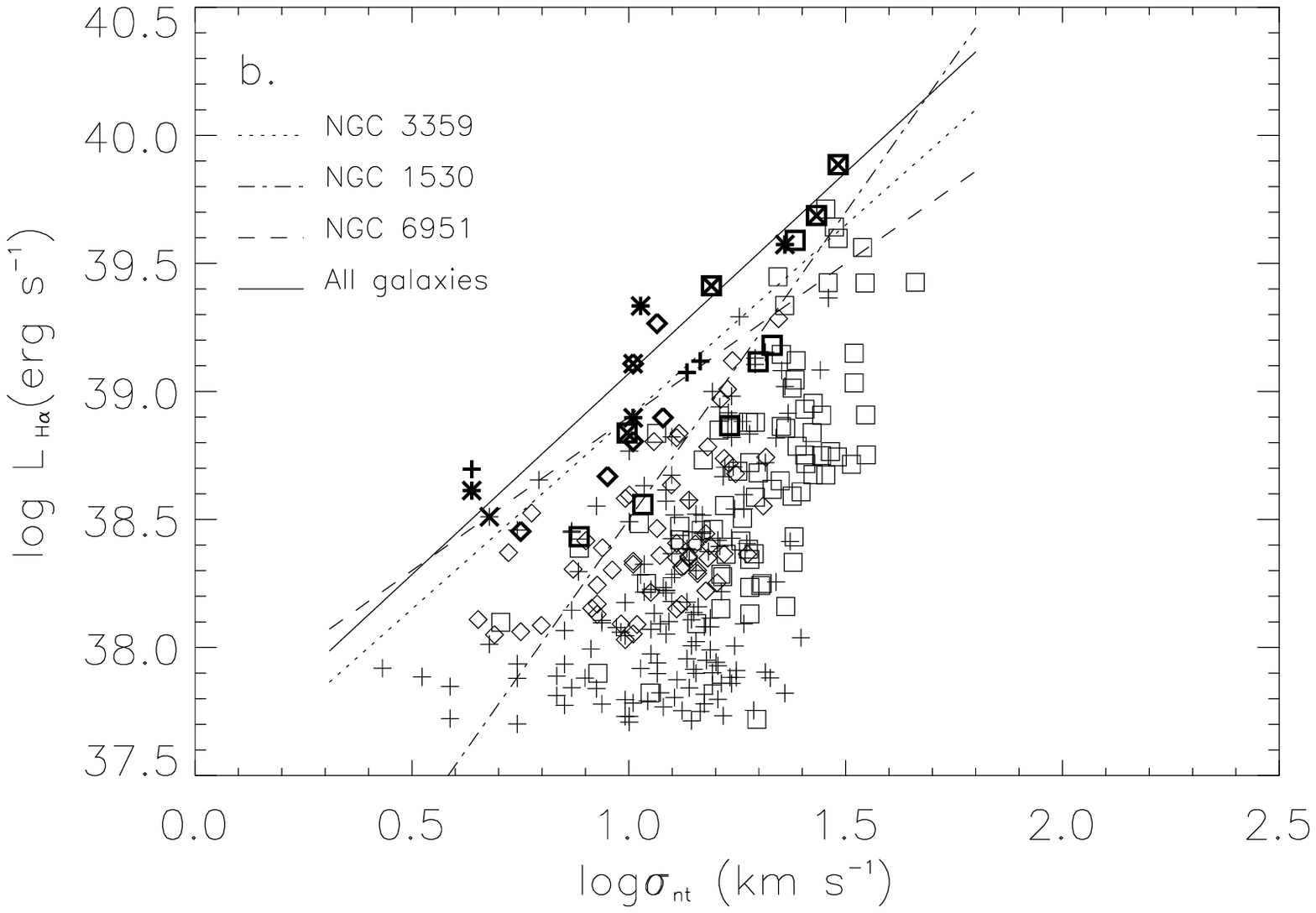}
\protect\caption[ ]{log~L$_{\scriptsize\ha}$--log~$\rm\sigma_{nt}$ diagram for the \hii\ regions in NGC~1530,
    NGC~3359 and NGC~6951 and the corresponding points defining 
    the best envelopes for each galaxy. Also shown is the representative linear fit to the
    envelope for each galaxy. The points marked with a cross represent the envelope when the \hii\ regions 
    of the three galaxies are considered as a single log~L$_{\scriptsize\ha}$--log~$\rm\sigma_{nt}$ distribution.
    a. the luminosity of the \hii\ region is that obtained from the catalogue, b. the 
    \ha\ luminosity of the \hii\ region is a fraction of the luminosity from the catalogue, 
    corresponding to the most intense Gaussian peak in the emission line profile.}
\label{Lsigmatresenv}
\end{figure*}

We have devised an automatic method to find an envelope for the
Figs.~\ref{Lsigmatres}a and~\ref{Lsigmatres}b. Starting at
a certain logarithmic \ha\ luminosity and increasing it in bins of fixed widths in logarithm of
luminosity, we selected for each bin the \hii\ region with the lowest
value of the non--thermal velocity dispersion. Changing the starting point in 
luminosity and the bin widths we optimized the envelope that best defines the 
log~L$_{\scriptsize\ha}$--log~$\rm\sigma_{nt}$
distribution. The points which best define the envelopes in the 
log~L$_{\scriptsize\ha}$--log~$\rm\sigma_{nt}$ diagrams for 
each galaxy and the distributions taking into account all the galaxies are shown in 
Figs.~\ref{Lsigmatresenv}a and~\ref{Lsigmatresenv}b. Although the lower limits in 
logarithmic \ha\ luminosity change from one galaxy to another, the bin widths that best 
define the envelope take the same value of 0.15 in all the distributions (see Tables~\ref{tabenvol} 
and~\ref{tabenvolfrac}). 

The best way of fitting the points on the envelope must take into account the uncertainties
in the logarithmic \ha\ lu\-mi\-no\-si\-ty of the \hii\ regions and in the velocity dispersion of the line profiles.
Except for 4 \hii\ regions on the envelope defined by NGC~6951, the rest of the
regions on the envelopes have luminosities above ${\rm
  log~L_{\scriptsize\ha}}>38.5$~(\ergs). The uncertainties
in \ha\ luminosity for \hii\ regions with a ${\rm
  log~L_{\scriptsize\ha}}>38.5$~(\ergs) is less than 10\% for NGC~3359, $\sim$10\% for
NGC~6951 and $\sim$20\% for NGC~1530. The high value for the uncertainty in
luminosity for the \hii\ regions in NGC~1530 is because this galaxy is further away than
the others and the image was taken with the JKT (1m Telescope); \hii\ regions with the same \ha\ luminosity but located at greater
distance will be seen with lower S:N ratio. The uncertainty in the velocity dispersion
is obtained from the estimated error given by the fit program in
de\-com\-po\-sing the line profiles. 

In order to obtain simultaneously 
both uncertainties we use a linear fit which gives a weight, w, to each point of the envelope defined by:

\begin{equation}
{\rm w}=((\Delta{\rm log~L_{\scriptsize\ha}})^2+(\Delta {\rm log~\rm\sigma_{nt}})^2)^{-1/2}
\end{equation}
where $\Delta{\rm log~L_{\scriptsize\ha}}$ and $\Delta{\rm log~\rm\sigma_{nt}}$ are the relative errors in
logarithmic \ha\ luminosity and logarithmic $\rm\sigma_{nt}$. 

We have taken into account the influence of the selected bin width and the logarithmic 
\ha\ luminosity starting point on the definition of the envelope, and specially on the slope of the envelope. 
In Table~\ref{tabbin}, we show the slopes for different selected bin widths. The values of the slopes 
are the median va\-lues of the slopes for the different starting points in the range of logarithmic \ha\ lu\-mi\-no\-si\-ty, and the
corresponding errors are the standard deviations for all the slopes for different starting points. As
can be seen from this table, a change in the bin width or in the starting point does not affect 
significantly the values of the slopes, which means that the procedure we have used is robust against these changes. 

\begin{table*}
\centering
\caption[]{Variation of the log~L$_{\scriptsize\ha}$-log~$\rm\sigma_{nt}$ envelope slope with the 
selected bin width. Each value and its error correspond respectively, to the median and standard 
deviation of the slopes for different bin starting points in log~L$_{\scriptsize\ha}$.}
\begin{tabular}{ccc|cc|cc}
\hline
\hline
bin width      & \multicolumn{2}{c}{NGC~1530}  & \multicolumn{2}{c}{NGC~6951} & \multicolumn{2}{c}{NGC~3359}   \\
\hline
    dex     & frac of L$_{\scriptsize \ha}$ & total L$_{\scriptsize \ha}$ & frac of L$_{\scriptsize \ha}$ & 
	 total L$_{\scriptsize \ha}$ & frac of L$_{\scriptsize \ha}$ & total L$_{\scriptsize \ha}$ \\ 
\hline
  0.10   & $2.4\pm0.2$ & $1.9\pm0.3$ & $0.8\pm0.2$& $0.8\pm0.2$& $1.6\pm0.3$& $1.1\pm0.3$ \\
  0.15   & $2.4\pm0.2$ & $2.0\pm0.2$ & $1.2\pm0.4$& $1.2\pm0.4$& $1.5\pm0.2$& $1.4\pm0.3$ \\
  0.20   & $2.4\pm0.2$ & $1.9\pm0.1$ & $1.6\pm1.9$& $1.6\pm1.9$& $1.6\pm0.2$& $1.4\pm0.3$ \\
\hline
\end{tabular}
\label{tabbin}
\end{table*}

Finally, we have selected as the representative fit to the envelope of each distribution that found 
from the variation of the starting point with a selected bin of 0.15, for which the envelopes 
are best defined. The $y$ coordinate of the zero point 
and the slope for each representative linear fit are the median va\-lues of the corresponding quantities, 
taken as the starting point in luminosity is varied. The results are shown in 
Tables~\ref{tabenvol} and~\ref{tabenvolfrac} and plotted in Figs.~\ref{Lsigmatresenv}a 
and~\ref{Lsigmatresenv}b. 

From Tables~\ref{tabenvol} and~\ref{tabenvolfrac} and the corresponding
Figs.~\ref{Lsigmatresenv}a and~\ref{Lsigmatresenv}b, we can see that
the envelope in the log~L$_{\scriptsize\ha}$--log~$\rm\sigma_{nt}$ distribution is not particularly well defined. The
envelopes for each galaxy have slopes with similar values but the errors in
the fits are large. In any case, the results of the linear fits for each galaxy agree
broadly with
the result of Arsenault \etal (1990). These authors found a limit in the upper boundary 
of ${\rm log~F_{\scriptsize\ha}}<(2.18~{\rm log~\rm\sigma_{nt}}-16.15)$ in the plane \ha\ surface
brightness - non--thermal velocity dispersion for the \hii\ regions they detected in
NGC~4321. 

\begin{table*}
\centering
\caption[]{Representative fits to the envelopes of the log~L$_{\scriptsize\ha}$--log~$\rm\sigma_{nt}$ distributions 
        for NGC~1530, NGC~3359 and NGC~6951 and for the distribution defined
        with all the \hii\ regions in the three galaxies. The logarithmic \ha\ 
        luminosity corresponds to the value given in the calibration catalogue.}
\begin{tabular}{lcc}
\hline
\hline
             & Bin & Envelope fit   \\
\hline
    NGC~1530 & 0.15 & log~L$_{\scriptsize\ha}$$\sim$(37.0$\pm$0.3)+(2.0$\pm$0.2)~log~$\rm\sigma_{nt}$   \\
    NGC~3359 & 0.15 & log~L$_{\scriptsize\ha}$$\sim$(37.5$\pm$0.4)+(1.4$\pm$0.7)~log~$\rm\sigma_{nt}$   \\
    NGC~6951 & 0.15 & log~L$_{\scriptsize\ha}$$\sim$(37.7$\pm$0.4)+(1.2$\pm$0.4)~log~$\rm\sigma_{nt}$   \\
All galaxies & 0.15 & log~L$_{\scriptsize\ha}$$\sim$(37.4$\pm$0.2)+(1.78$\pm$0.14)~log~$\rm\sigma_{nt}$   \\
\hline
\end{tabular}
\label{tabenvol}
\end{table*}

\begin{table*}
\centering
\caption[]{The same parameters as shown in Table~\ref{tabenvol}, but the logarithmic \ha\
        luminosity is given by the fraction of the logarithmic \ha\ luminosity of the
        catalogue corresponding to the most intense Gaussian peak.}
\begin{tabular}{lcc}
\hline
\hline
             & Bin & Envelope fit   \\
             \hline
     NGC~1530 & 0.15 & log~L$_{\scriptsize\ha}$$\sim$(36.1$\pm$0.2)+(2.4$\pm$0.2)~log~$\rm\sigma_{nt}$ \\
     NGC~3359 & 0.15 & log~L$_{\scriptsize\ha}$$\sim$(37.4$\pm$0.2)+(1.5$\pm$0.7)~log~$\rm\sigma_{nt}$ \\
     NGC~6951 & 0.15 & log~L$_{\scriptsize\ha}$$\sim$(37.7$\pm$0.4)+(1.2$\pm$0.4)~log~$\rm\sigma_{nt}$ \\
 All galaxies & 0.15 & log~L$_{\scriptsize\ha}$$\sim$(37.5$\pm$0.1)+(1.57$\pm$0.07)~log~$\rm\sigma_{nt}$ \\            
\hline  
\end{tabular}
\label{tabenvolfrac}
\end{table*}

\subsubsection{Envelope for the best determined velocity dispersion}

In order to find better defined
envelopes than those shown in Figs.~\ref{Lsigmatresenv}a and~\ref{Lsigmatresenv}b and minimize the errors in the slopes of the
envelopes, we have eliminated the \hii\ regions in the log~L$_{\scriptsize\ha}$--log~$\rm\sigma_{nt}$ diagram which have
large errors in their velocity dispersions. From Table~\ref{errLsigma} we see 
that the regions with $\rm\sigma_{nt}<13$~\kms\ have relative errors of $\sim$28\%, so we have
eliminated them from the log~L$_{\scriptsize\ha}$--log~$\rm\sigma_{nt}$ diagram and extracted again the envelope for
all the galaxies. 

\begin{figure}
\centering
\includegraphics[width=9cm]{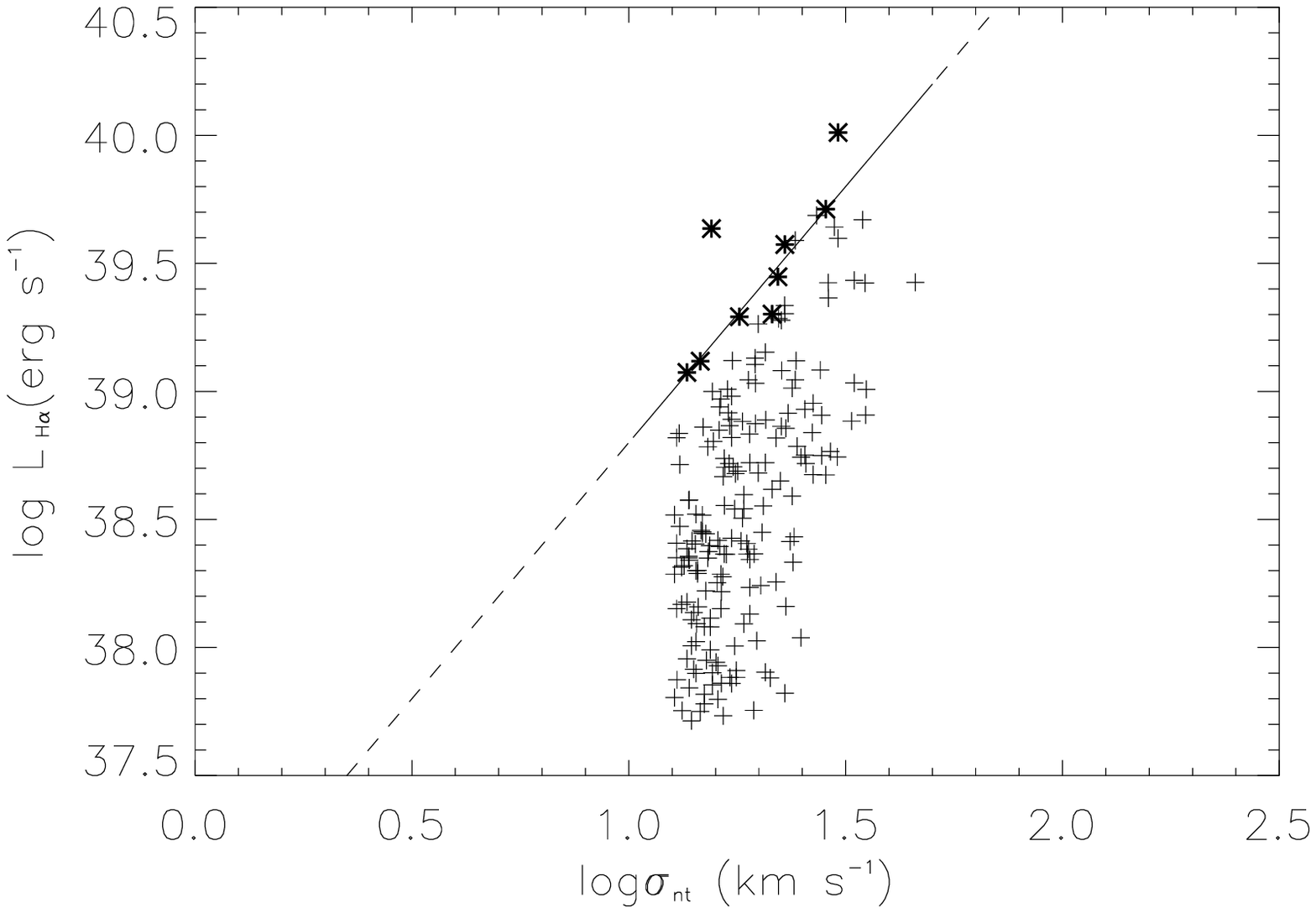}
\protect\caption[ ]{log~L$_{\scriptsize\ha}$--log~$\rm\sigma_{nt}$ diagram for the three galaxies with \hii\
  regions having $\rm\sigma_{nt}>13$~\kms. The continous line represents the linear fit to the points on the
  envelope (bolder points in the diagram). The logarithmic \ha\
  luminosities are those in the photometric \hii\ region catalogue.}
\label{Lsigmacorte}
\end{figure}

\begin{figure}
\centering
\includegraphics[width=9cm]{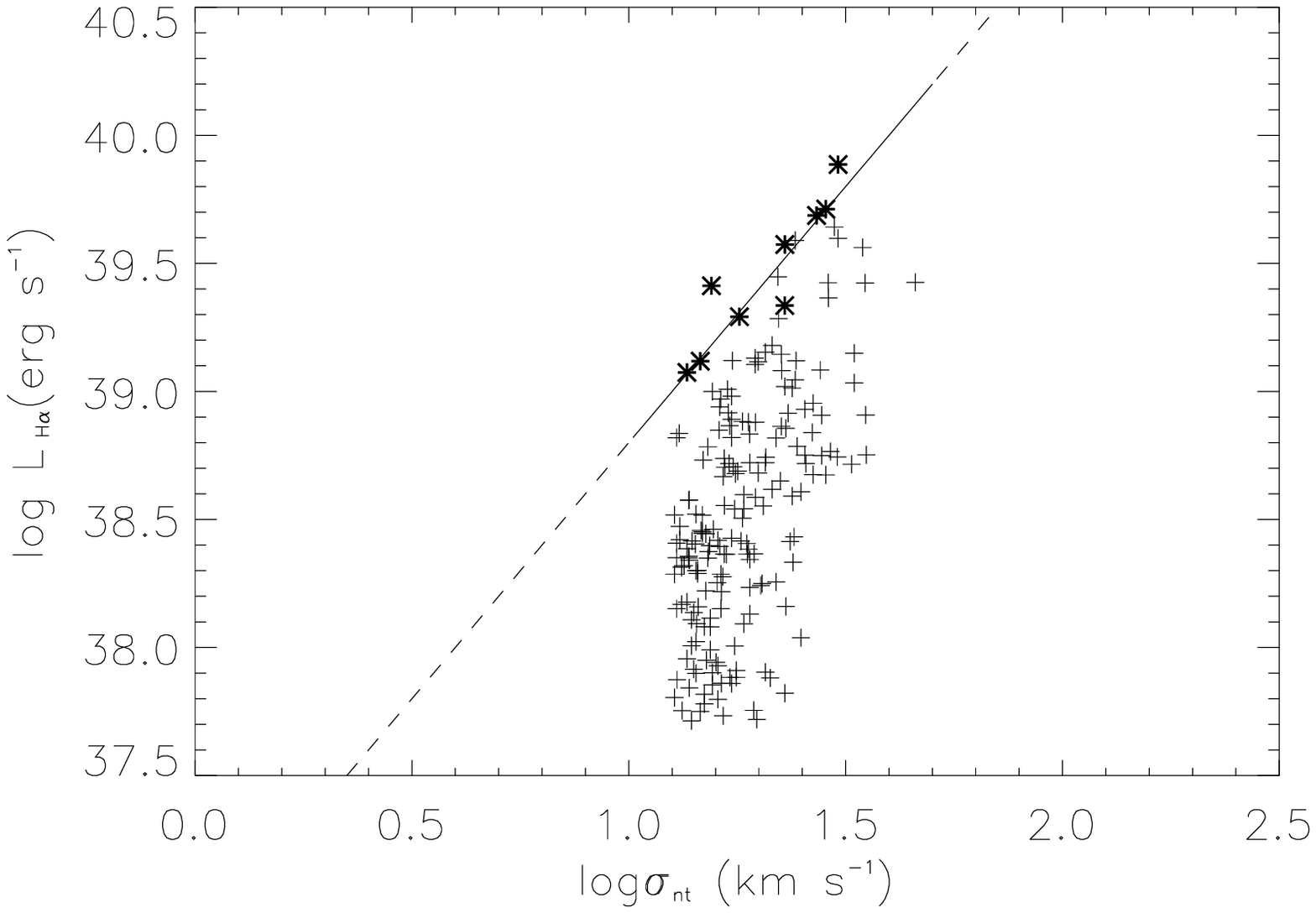}
\protect\caption[ ]{log~L$_{\scriptsize\ha}$--log~$\rm\sigma_{nt}$ diagram for the three galaxies with \hii\
  regions having $\rm\sigma_{nt}>13$~\kms\ as in Fig.~\ref{Lsigmacorte}, but the 
  logarithmic \ha\ luminosities of the \hii\ regions are found as a fraction of the logarithmic \ha\ 
  luminosity from the catalogue, as explained in text. The envelope fit is somewhat tighter than that
  in Fig.~\ref{Lsigmacorte}, but the slope is the same.}
\label{Lsigmacortefrac}
\end{figure}

The results are
shown in Figs.~\ref{Lsigmacorte} and~\ref{Lsigmacortefrac}. More clearly defined envelopes are defined
for the \hii\ regions with $\rm\sigma_{nt}>13$~\kms, for both distributions: when
the total logarithmic \ha\ luminosity from the catalogue is considered and when an appropiate
fraction is used for those \hii\ regions with the central peak defined by two
Gaussian components of comparable intensities. The linear fit for the envelope is 
${\rm log~L_{\scriptsize\ha}}=(36.8\pm0.7)+(2.0\pm0.5)~{\rm log~\rm\sigma_{nt}}$ in the case where the logarithmic \ha\ luminosity of 
the \hii\ region is given by the catalogue (Fig.~\ref{Lsigmacorte}). For the case 
where the logarithmic \ha\ lu\-mi\-no\-si\-ty is a fraction of the \hii\ region luminosity, the linear fit is
${\rm log~L_{\scriptsize\ha}}=(36.8\pm0.6)+(2.0\pm0.5)~{\rm log~\rm\sigma_{nt}}$ (see Fig.~\ref{Lsigmacortefrac}). 

Although the results found here do not agree with the L--$\sigma$ relation proposed by 
Terlevich \& Melnick (1981), they do 
agree quite well with those found by Arsenault \etal (1990) for the \hii\ regions with the highest surface
brightness in NGC~4321. They found that these regions can be fitted by
${\rm log~F}_{\scriptsize\ha}=(2.57\pm0.49)~{\rm log~\rm\sigma_{nt}}-(17.21\pm0.39)$, 
where ${\rm log~F}_{\scriptsize\ha}$ is the surface brightness of the \hii\ region.
Rozas \etal (1998) found a linear fit to the \hii\ regions on the envelope of ${\rm
  log~L_{\scriptsize\ha}}=36.15+2.6~{\rm log~\rm\sigma_{nt}}$
for the \hii\ regions in NGC~4321 with central component widths above
10~\kms, and this linear fit is in fair agreement with those shown in Tables~\ref{tabenvol}
and~\ref{tabenvolfrac} and also with the linear fits obtained for \hii\ regions 
with $\rm\sigma_{nt}>13$~\kms. 

Finally, it is also interesting to note that the slopes of the envelopes for both
log~L$_{\scriptsize\ha}$--log~$\rm\sigma_{nt}$ diagrams obtained by ta\-king
the luminosity of an \hii\ region directly from the catalogue, and
the diagram taking the logarithmic \ha\ luminosity of the region as a fraction of the
catalogued luminosity, are the same. However, the log~L$_{\scriptsize\ha}$--log~$\rm\sigma_{nt}$ diagram taking 
into account the fraction of the logarithmic \ha\ luminosity gives an envelope with less
dispersion in its defining points. 

\subsubsection{Testing for metallicity dependence}

The metallicity could affect the log~L$_{\scriptsize\ha}$-log~$\rm\sigma_{nt}$ relationship as it affects the
thermal velocity dispersion of a given region
because the electron temperature of the \hii\ region depends on the metallicity 
and in this way the metallicity gradient across the disc of the spiral
galaxies could be
affecting the log~L$_{\scriptsize\ha}$--log~$\rm\sigma_{nt}$ relation, as was first proposed by Terlevich \& Melnick (1981).
We check here this posibility.

We have assumed a constant electron temperature of $10^4$~K for all the \hii\ 
regions, which gives a thermal broadening of 9.1~\kms. Electron tem\-pe\-ra\-tu\-res
of 5000~K and 15000~K give thermal broadenings of 6.4~\kms and 11.1~\kms
respectively. Assuming a typical observed velocity dispersion of
$\sigma=(22.0\pm2.0)$~\kms\ and a $\sigma_{\rm inst}=(16.3\pm0.3)$~\kms, the
non--thermal velocity dispersions for different temperatures are 
$\sigma_{\rm nt}(\rm T=5000~K)=(13.0\pm3.8)$~\kms, 
$\sigma_{\rm nt}(\rm T=10000~K)=(11.2\pm4.3)$~\kms and $\sigma_{\rm nt}(\rm
T=15000~K)=(9.3\pm5.3)$~\kms. Thus, the assumption of a constant temperature
of $10^4$~K for all the \hii\ regions is a very reasonable approximation and the effect on 
the non--thermal velocity dispersion for a reasonable spread on the temperatures will be
inside the error bars.

\begin{figure}
\centering
\includegraphics[width=9cm]{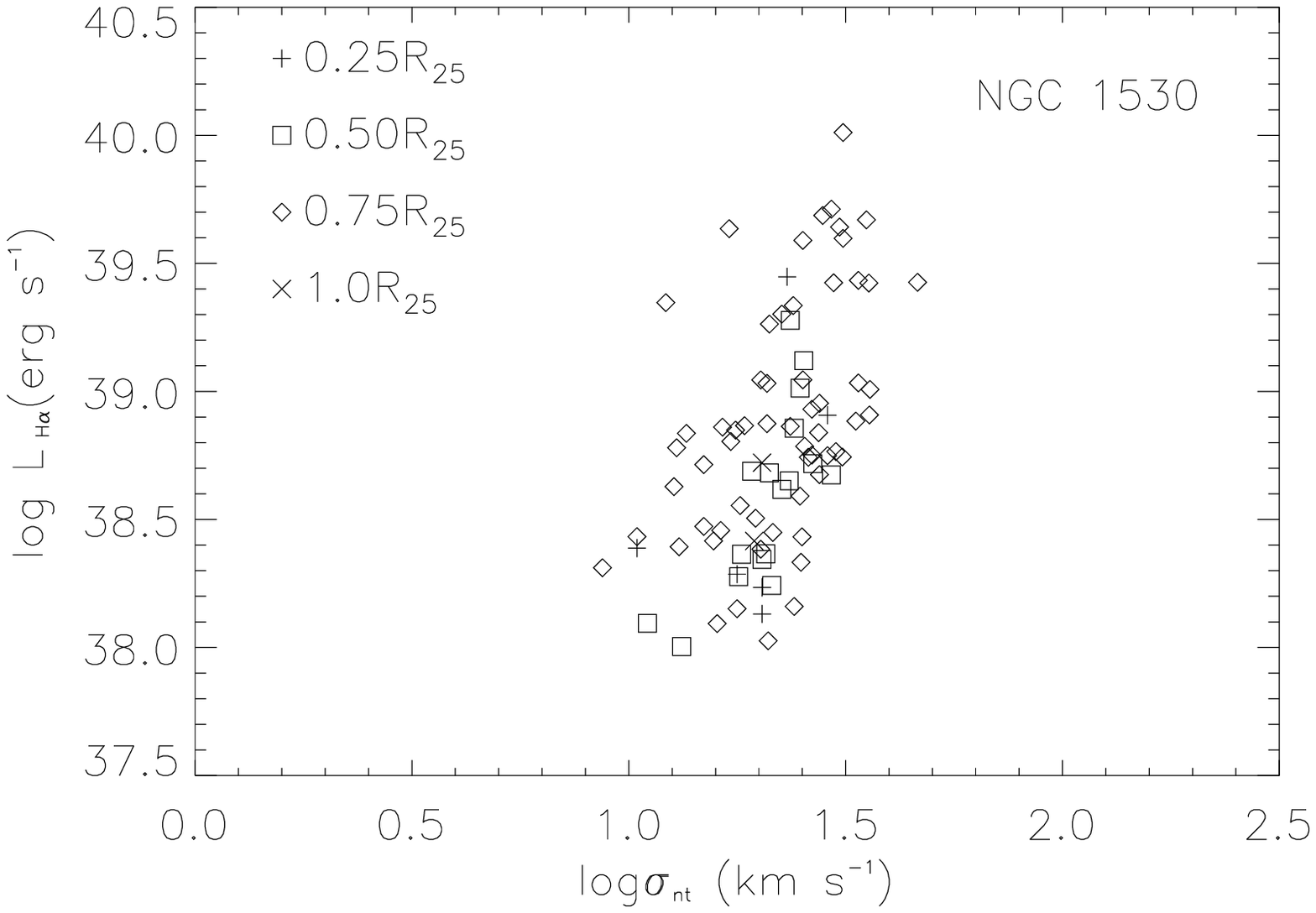}
\includegraphics[width=9cm]{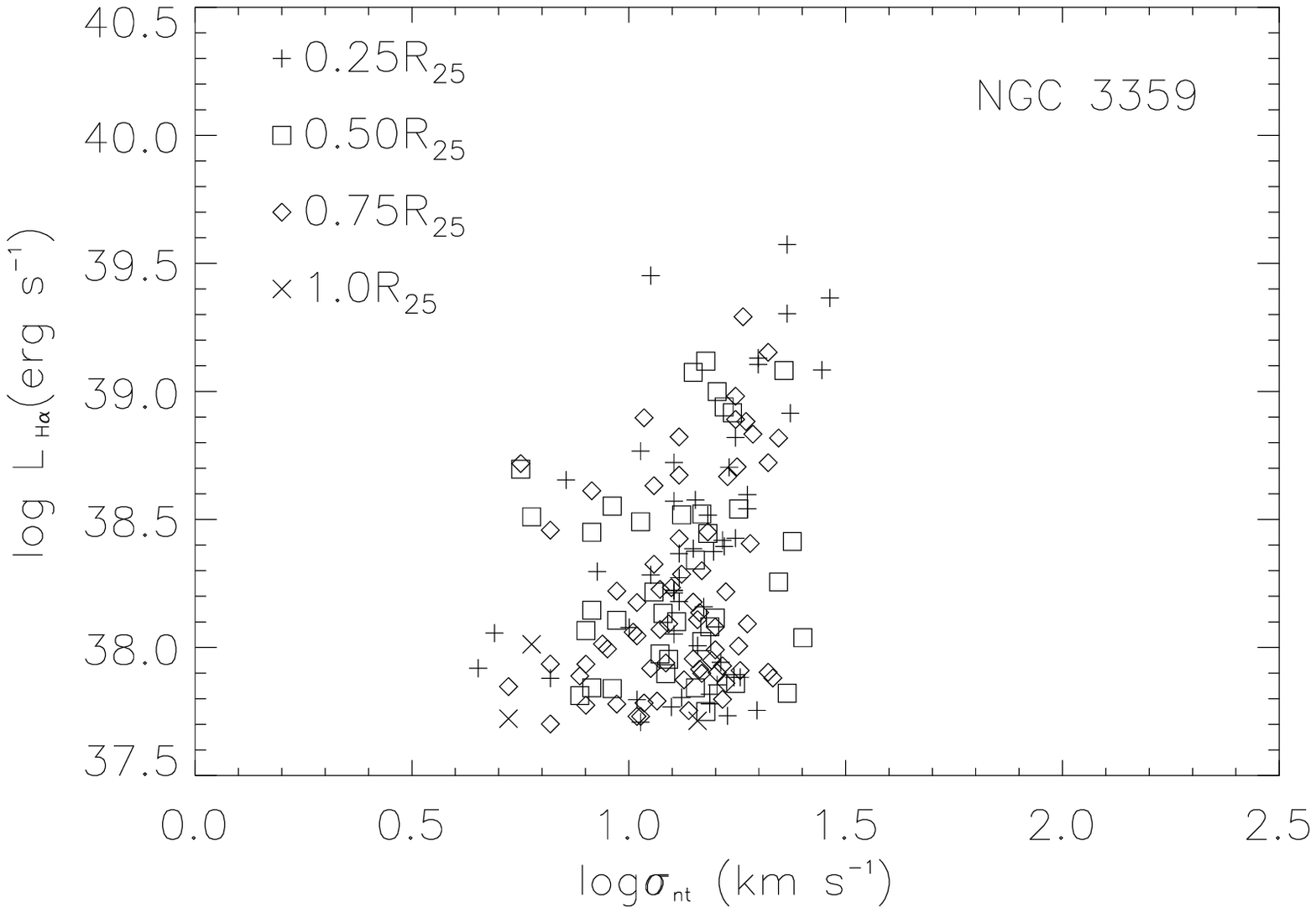}
\includegraphics[width=9cm]{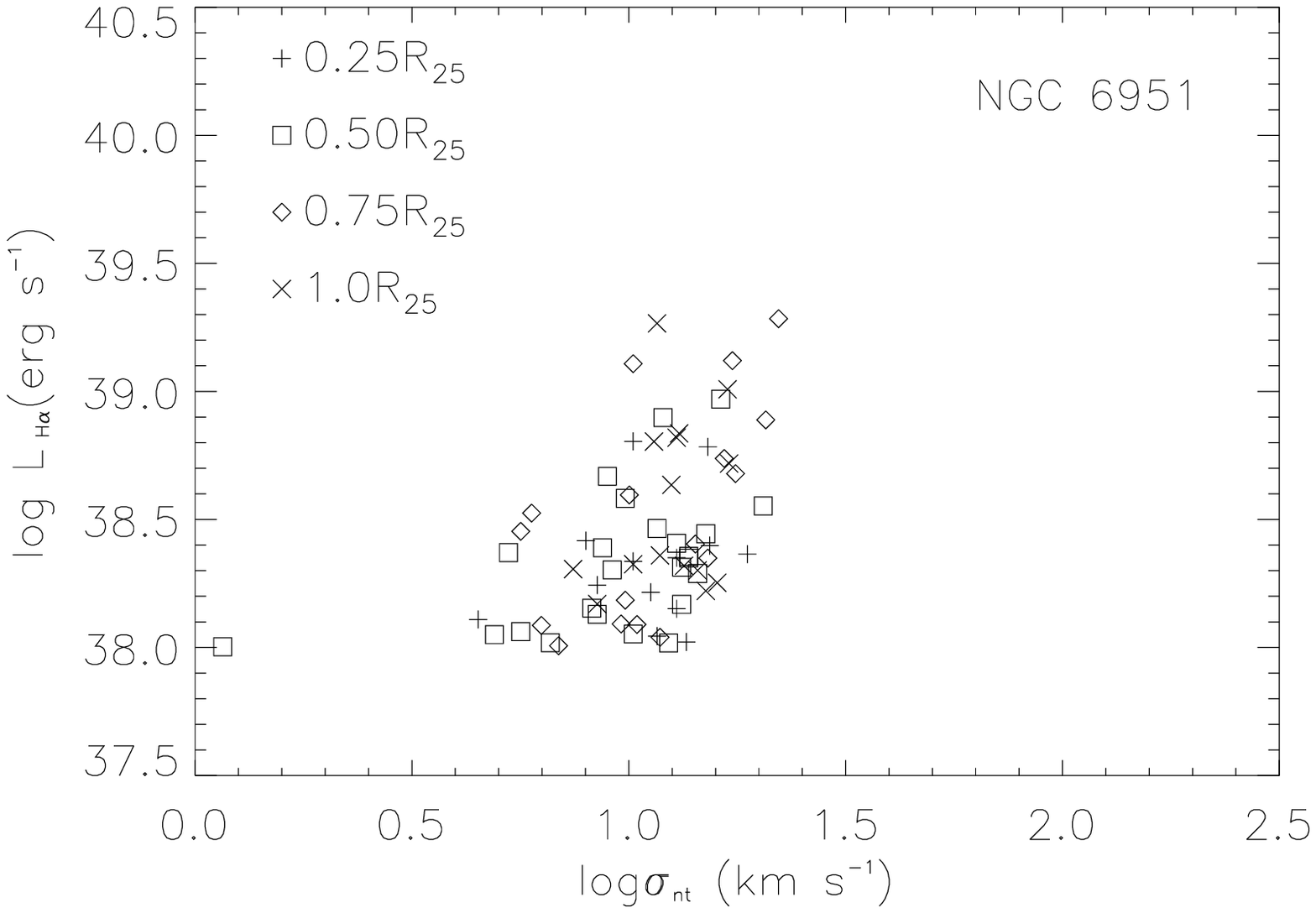}
\protect\caption[ ]{Plot of the \hii\ regions at different deprojected galactocentric
  radii in the log~L$_{\scriptsize\ha}$--log~$\rm\sigma_{nt}$ diagram for the three ga\-la\-xies NGC~1530, NGC~3359 
  and NGC~6951, showing that radial effects (and hence by implication,
  metallicity effects) are not affecting systematically the log~L$_{\scriptsize\ha}$--log~$\rm\sigma_{nt}$ relation.}
\label{Lsigmetalicidad}
\end{figure}

A test for how the radial metallicity gradient might
be affecting the log~L$_{\scriptsize\ha}$--log~$\rm\sigma_{nt}$ relation is shown
in Fig.~\ref{Lsigmetalicidad}, where we show the positions in the
log~L$_{\scriptsize\ha}$--log~$\rm\sigma_{nt}$ diagram of the \hii\ regions of each galaxy located in 
increasing ranges of galactocentric radius. We take the length of the major axis of the galaxy out to a 
a blue surface brightness level of 25~mag~arcsec$^{-2}$, $\rm R_{25}$, as the radial unit. 
The values of $\rm R_{25}$ for each galaxy are taken
from de Vaucouleurs \etal (1991). 
From Fig.~\ref{Lsigmetalicidad} it can be seen that there is no trend of the
\hii\ regions at a given galactocentric radius, which presumably all have essentially the same
metallicity, to be concentrated in a certain zone of the log~L$_{\scriptsize\ha}$--log~$\rm\sigma_{nt}$ diagram.
There is no observed sys\-te\-ma\-tic trend of $\rm\sigma_{nt}$ with galactocentric radius
for a given lu\-mi\-no\-si\-ty, so we can discount the importance of the metallicity
in this context.

\section{The masses of the H~II regions in the log~L$_{\scriptsize\ha}$--log~$\rm\sigma_{nt}$ diagram}
\label{secmasasviri}

\subsection{Virial Mass}

The virial mass is obtained from Eq.~\ref{masviri} resulting from
the application of the virial theorem.

\begin{equation}
\label{masviri}
\rm M=\frac{\rm\sigma_{nt}^{2}R}{G}
\end{equation}
where $G$ is the gravitational constant, R is a specified radius of the \hii\ region, 
and $\rm\sigma_{nt}$ is the non-thermal velocity dispersion of the line profile, which 
is taken to be a representative value of the velocity of the gas within the 
region. 

A good estimate of the relevant
radius is subject to uncertainties which in most cases do
not allow us to obtain a fiducial value. In order to obtain the virial
mass, we have taken an estimate of the radius which is half of the
radius in the \hii\ region catalogues. This is consistent with the method of
Arsenault \etal (1990), who take the radius containing 40\% of the total flux 
of the \hii\ regions to characterize their sizes and extract the line
profiles with an aperture having twice these characteristic va\-lues. They based this 
criterion on the study of McCall, Hill \& English (1990), who show that the 40\% isophote is a 
consistent measure of the \hii\ region size because it is independent of the 
detection threshold and the calibration. The virial mass computed in this way 
for the \hii\ regions on the envelope and a selection of
regions well below the envelope in luminosity in the log~L$_{\scriptsize\ha}$--log~$\rm\sigma_{nt}$ diagram is shown in
Tables~\ref{Tabmassenv} and~\ref{Tabmass} respectively.

\subsection{Mass of the H~II region inferred from the \ha\ luminosity}

The mass of the gas within the region is obtained by integrating over 
the measured volume of the region and multiplying it by the mass of the hydrogen 
atom, thus:

\begin{equation}
\label{mastot}
{\rm M_{reg}}=f\int \phi^{1/2}<{\rm N}_{\rm e}>_{\rm rms}{\rm m}_{\rm p}{\rm dV}
\end{equation}
where f is a factor accoun\-ting for the fraction of the neutral mass inside the
region, $\phi$ is the filling factor and ${\rm <N_{e}>_{rms}}$ is the root mean square (r.m.s) electron
density. The factor $f$, accounting for the fraction of the neutral mass inside the
region that cannot be detected by the \ha\ observations, is estimated to take a value 
$\sim10$ from models by Giammanco \etal (2004).

In that paper, a model has been proposed which gives a good 
account of the  propagation of ionizing radiation in \hii\ regions making well 
defined assumptions  about the structure of the regions, and using these to compare with 
their observed properties. One of its key results is that the major 
fraction of the gas in an \hii\ region (and this applies to regions of all sizes, though 
the largest regions may show a somewhat smaller effect than the smallest) will remain 
neutral due to the pronounced inhomogeneity which is a well attested feature of the density 
distribution. In the paper a range of estimates of the neutral mass fraction inside the \hii\ 
regions is quoted, with a canonical value of close to an order of magnitude.

The r.m.s. electron density, ${\rm <N_{e}>_{rms}=\sqrt{<N_{e}^{2}>}}$, can be estimated
from the emission measure (EM) of the \hii\ region. The Emission
Measure is defined as the square of the electron density integrated along the 
line of sight.

\begin{equation}
\label{EMdef}
{\rm EM=\int_{0}^{s} N_{e}^{2}ds \approx <N_{e}^{2}><s>}
\end{equation}
where $\rm <s>$ is the average value of the line of sight length across the \hii\
region over the projected area of the region, (see Zurita 2001). 
The emission measure can be related to the surface brightness of the \hii\ region via the
theoretical formula given by Spitzer (1978):

\begin{equation}
\label{brillo}
{\rm {L_{\scriptsize\ha} \over \pi R^2} = h\nu_{\scriptsize\ha} 
\alpha_{\scriptsize\ha}^{eff}(H_{0},T) {N_{p} \over  N_{e}} 2.46\times 10^{17}\times {\rm EM}} 
\end{equation}
where ${\rm \alpha_{\scriptsize\ha}^{eff}(H_{0},T)}$ is the effective
recombination coefficient of the \ha\ emission line, ${\rm h\nu_{\scriptsize\ha}}$ 
is the energy of an \ha\ photon and R 
is the radius (in cm) of the \hii\ region. From the \ha\ luminosity and 
the radius obtained from the catalogues we can obtain the EM using Eq.~\ref{brillo} and the ${\rm
  <N_{e}>_{rms}}$ using Eq.~\ref{EMdef}. Values for these quantities are shown in
  Tables~\ref{Tabmassenv} and~\ref{Tabmass}.

The filling factor $\phi$ is defined to take into account inhomegeneities in
the density of the \hii\ regions. The inhomogeneities are schematized as
condensations of gas, with electron density ${\rm N_{e}}$ within the condensations,
and with negligible electron density in the intervening volume. The
filling factor can be computed using the expression (see Appendix~D in Zurita (2001)):

\begin{equation}
\label{ff}
\phi={\rm (\frac{<N_{e}>_{rms}}{N_{e}})^{2}}
\end{equation}

The values of $\phi$ for a selected set of \hii\ regions are shown in Tables~\ref{Tabmassenv}
and~\ref{Tabmass}. We adopt the mean value of ${\rm N_{e}\sim
  135}$~\cmtres\ taken from measurements by Zaritsky (1994). With the values of ${\rm
  <N_{e}>_{rms}}$ and $\phi$, we obtain the total mass of the \hii\ region via 
  Eq.~\ref{mastot}; the results are shown in Tables~\ref{Tabmassenv} and~\ref{Tabmass}.

\subsubsection{Mass of the stellar content in the \hii\ region}

The total mass of a star cluster can be obtained by integrating the initial mass
function (IMF), which gives the number of stars formed at the same time in a
given range mass, over the total range mass ${\rm (m_{low},m_{up})}$. 

\begin{equation}
\label{IMF}
{\rm M_{stellar}=\int_{m_{low}}^{m_{up}} m\Phi(m)dm}
\end{equation}
where ${\rm \Phi(m)}$ is the IMF, normally assumed to be a power of m, ${\rm
  \Phi(m)=Am^{\alpha}}$, $\alpha=-2.35$ for the Salpeter IMF (Salpeter 1955) and A is a
normalisation factor. 

In order to estimate the mass of the stars within the
\hii\ region, we need to know A, the mass, ${\rm m_{low}}$, of the lowest mass star
and the mass, ${\rm m_{up}}$, of the highest mass
star. The nor\-ma\-li\-za\-tion factor A can be obtained from the
estimate of the mass of the ionizing stars, as we can put a lower physical
limit $\rm m_{l_{i}}$ on the mass of these stars which contribute significantly to the
ionization of the \hii\ region. This is done in the following way. 

The {\it ionizing mass} is calculated by
integrating over the masses of the stars that contribute significantly to the
ionizing radiation using:

\begin{equation}
\label{IMFioni}
{\rm M_{ionizing}=\int_{m_{l_{i}}}^{m_{up}} A m^{-1.35}dm}
\end{equation}
The \ha\ luminosity gives an equivalent number of O5(V) stars and using the
mass of an O5(V) spectral type star from Vacca \etal (1996), an estimate of the
ionizing mass inside the \hii\ region can be obtained. The mass limits in
Eq.~\ref{IMFioni} are found assu\-ming that the most massive ionizing star is
an O3(V) type star and that the sum of all the stars with spectral type O9(V) or later
do not contribute more than 10\% to the total ioni\-zing flux. Thus, 
${\rm m_{l_{i}}}=22.1\rm M_\odot$ and ${\rm m_{up}}=51.3\rm M_\odot$ from the values from Vacca \etal (1996)
for the masses of spectral type stars O3(V) and O9(V) respectively. 

Once the normalization factor is known, the mass for the stellar cluster is
obtained by integrating Eq.~\ref{IMF}. In Tables~\ref{Tabmassenv} and~\ref{Tabmass} we show
the total masses for two groups of \hii\ regions. The
integration limits are ${\rm (0.1\Msun,100\Msun)}$. Using this mass range gives an
upper limit to the stellar cluster mass, since in a real cluster the curve
should be less steep than the Salpeter function for masses less than $\rm 1\Msun$ .

The mass of the stellar content obtained in this way is meant as an estimate which
allows us to obtain an order of magnitude idea of the fraction of the \hii\ region mass that is
in stellar form. The Salpeter IMF is not necessarily the most representative functional form to
describe the stellar content of the cluster that produces an \hii\
region. Massey \etal (1989) obtained a slope for the IMF of $(-2.9\pm0.3)$ in
the 9-85~$\rm\Msun$ range for NGC~346 in SMC, and Rela\~no \etal (2002) found a slope of
$(-3.7\pm0.4)$ in the 24-54.1~$\rm\Msun$ range. In
addition, the total ioni\-zing ste\-llar mass obtained from the observed \ha\
luminosity of the \hii\ region does not take into account the escape fraction of luminosity in density
bounded regions and the equivalent number of stars 
of spectral type O5(V) gives only an approximation to the total ionizing
mass. In fact, the total mass obtained from the equivalent number of O5(V)
stars in NGC~346 represents 60\% of the mass for the stars from spectral
types O3(V) to O9(V) (see Rela\~no \etal 2002). 

These uncertainties do not allow us to find a very accurate value of the
mass of the stars within an \hii\ region, but we can obtain a fair estimate of the
fraction of the total \hii\
region mass in stellar form. Applying different mass limits in Eq.~\ref{IMF} gives a fraction 
of the total \hii\ region mass in stellar form of 0.17\% and 0.44\% for mass ranges ${\rm (1\Msun,100\Msun)}$ and ${\rm
  (0.1\Msun,100\Msun)}$, respectively. If the mass obtained from the
equivalent number of O5(V) spectral type stars were 60\% of the total mass
for stars earlier than O9(V), as is the case for NGC~346, the total
mass in stellar form would not be more than 1\% of the \hii\ region mass. Moreover,
even using a much steeper IMF, $\alpha=-3.7$ as found in NGC~346, and integrating
over a mass range of ${\rm (1\Msun,100\Msun)}$ gives a fraction
 of only 4.7\% of the region mass in stellar form. These estimates agree with
 the stellar mass content for NGC~604 given by Yang \etal (1996), who estimated
 a fraction of stellar mass content in the total mass of the region between 4\% and 1\%. 
 
 Using these considerations we can see that the mass of the stellar cluster represents
 a relatively small fraction of the \hii\ region mass and
 for this reason it may be neglected in the
 comparisons between the \hii\ region mass obtained from the \ha\ luminosity and 
 the virial mass. 

\subsection{Discussion}

As can be seen from Tables~\ref{Tabmassenv} and~\ref{Tabmass}, the virially estimated
mass of the \hii\ regions located on the envelope is close to the total
mass of the \hii\ region, the ratio M$_{\rm vir}$/M$_{\rm reg}$ is between 1 and 4 (last column in
Table~\ref{Tabmassenv}). Virial masses for the
\hii\ regions located well above the envelope in $\sigma_{\rm nt}$ are however bigger than their
corresponding total \hii\ region masses, the ratios M$_{\rm vir}$/M$_{\rm reg}$
range from 7 to 15 as shown in last column in Table~\ref{Tabmass}. This result shows that 
the \hii\ regions located on the envelope may well be virialized systems or at least
close to virial equilibrium, while the velocity dispersions from the \hii\ regions located well above the envelope in $\sigma_{\rm nt}$
in the log~L$_{\scriptsize\ha}$--log~$\rm\sigma_{nt}$ diagram show the presence of other processes that affect
the internal kinematics of the \hii\ regions.

The fact that the velocity dispersions of the principal velocity component can be affected 
by a number of processes has been investigated by Yang \etal (1996) for the nearby region NGC~604 in M33. 
They mapped this region with 
Fabry--P\'erot interferometry and extracted line profiles for each pixel in the map. The 
velocity dispersion distribution for all the positional line profiles show that the numerous 
points with high intensities define a narrow band with a mean value of 
$\rm\sigma_{nt}\sim$12~\kms, and a lower limit of $\rm\sigma_{nt}\sim$9~\kms, 
while the few low intensity points have higher velocity dispersions. From comparison of the 
total mass of NGC~604 and the virial mass obtained from the 
underlying velocity dispersion values they conclude that gravitation may provide the 
basic mechanism for most positions, while other mechanisms are responsible of the 
excess broadening at the low intensity points. It is clear that these mechanisms affect 
even the \hii\ regions close to the envelope in Fig.~\ref{Lsigmacortefrac}, but certainly 
affect even more strongly the ma\-jo\-ri\-ty for the \hii\ regions in our sample. They include 
specifically stellar winds (e.g. Dyson (1979)), supernova explosions 
(e. g. Skillman \& Balick (1984)) and the cumulative kinematic effects of these 
have been termed champagne flows (e. g. Tenorio-Tagle (1979)).

The integrated line profile for NGC~604 that Yang \etal (1996) obtained, 
presents a central component with $\rm\sigma_{nt}\sim$15~\kms, higher than the 
dominant value for the underlying velocity dispersion, $\rm\sigma_{nt}\sim$12~\kms. This 
means that in the width of the integrated line profile of the \hii\ region there is 
an unavoidable contamination by the lower intensity points with higher widths.

Following this reasoning and taking $\rm\sigma_{nt}\sim$10.5~\kms\ (the mean of    
$\rm\sigma_{nt}\sim$9~\kms\ and $\rm\sigma_{nt}\sim$12~\kms, which define the range 
in velocity dispersion of the most numerous high intensity points in NGC~604, 
we expect that a fraction of $10.5/15=0.7$ of the non--thermal 
velocity dispersions of the integrated line profiles are due to contamination by other 
mechanisms than gravitation. If this is so, a corrected value of  
$\rm\sigma_{cor}=0.7\sigma_{nt}$ in Eq.~\ref{masviri} would make a correction in 
the virial mass of $\rm M_{cor}=0.5M_{vir}$, which gives ratios of $\rm M_{cor}/M_{reg}\sim 1.5$.

The uncertainties we have shown in our estimates of the correct value of the non--thermal velocity 
dispersion and of the relevant radius to use in Eq.~\ref{masviri}, plus 
the result $\rm M_{cor}/M_{reg}\sim 1.5$ 
lead us to conclude that the masses of the \hii\ regions 
located on the envelope are in fact consistent with masses obtained using the virial 
equilibrium equation.

\begin{table*}
\centering
\caption[]{Logarithmic \ha\ luminosity, emission measure, r.m.s. electron density, filling factor, virial
  mass, total mass of the \hii\ region from the \ha\ emission, mass of the stellar content in the
  \hii\ region and the fraction of the mass of the region that
  represents the virial mass for the \hii\ regions located on the
  envelope of Fig.~\ref{Lsigmacorte} (without taking non--virial contamination into account, see text). 
}
\begin{tabular}{ccccccccc}
\hline
\hline
Region & log~L$_{\scriptsize\ha}$ & EM & ${\rm <N_{\rm e}>_{rms}}$ & $\phi$ & M$_{\rm vir}$ &
M$_{\rm reg}$ & M$_{\rm stellar}$ & M$_{\rm vir}$/M$_{\rm reg}$ \\ 
       &(\ergs) & (pc~\cmseis) & (\cmtres) & ($10^{-4}$) & ($10^{6}\rm \Msun$) &
       ($10^{6}\rm\Msun$) & ($10^{4}\rm\Msun$) & \\
\hline
13(N3359)  &   39.07 &   4500 & 3.24 &  5.8   &  6.9  &  2.6 & 2.3  & 2.63 \\
9(N3359)   &   39.12 &   7400 & 4.59 &  0.12  &  6.5  &  2.9 & 2.5  & 2.24 \\
5(N3359)   &   39.29 &   5800 & 3.48 &  6.6   &  13.6 &  4.3 & 3.8  & 3.13 \\
16(N1530)  &   39.30 &  10000 & 5.23 &  0.20  &  14.7 &  4.4 & 3.9  & 3.32 \\
8(N1530)   &   39.45 &   6400 & 3.42 &  6.4   &  23.2 &  6.2 & 5.4  & 3.75 \\
1(N3359)   &   39.57 &   6000 & 3.04 &  5.1   &  29.8 &  8.3 & 7.2  & 3.60 \\
5(N1530)   &   39.64 &  14500 & 5.67 &  0.20  &  9.5  &  9.6 & 8.4  & 0.99 \\
2(N1530)   &   39.71 &  10000 & 4.13 &  9.3   &  41.7 & 11.4 & 10.0 & 3.66 \\
1(N1530)   &   40.01 &  15000 & 4.66 &  0.11  &  55.3 & 22.7 & 19.8 & 2.44 \\
\hline
\end{tabular}
\label{Tabmassenv}
\end{table*}

\begin{table*}
\centering
\caption[]{Same parameters as in Table~\ref{Tabmassenv} but for a group of \hii\ regions located 
well above the envelope in $\sigma_{\rm nt}$ of Fig.~\ref{Lsigmacorte}.}
\begin{tabular}{ccccccccc}
\hline
\hline
Region & log~L$_{\scriptsize\ha}$ & EM & ${\rm <N_{e}>_{rms}}$ & $\phi$ & M$_{\rm vir}$ &
M$_{\rm reg}$ & M$_{\rm stellar}$ & M$_{\rm vir}$/M$_{\rm reg}$ \\ 
       & (\ergs) & (pc~\cmseis) & (\cmtres) & ($10^{-4}$) & ($10^{6}\rm\Msun$) & ($10^{6}\rm\Msun$) & ($10^{4}\rm\Msun$) & \\
\hline
26(N1530)  &    39.03 & 6700 &  4.48 & 0.11 & 31.7 & 2.4 & 2.1 &  13.31 \\
33(N1530)  &    38.86 & 4800 &  3.84 &  8.1 & 14.3 & 1.6 & 1.4 &   8.84 \\
36(N1530)  &    38.88 & 6000 &  4.43 &  0.1 & 27.7 & 1.7 & 1.5 &  16.38 \\
46(N1530)  &    38.79 & 5000 &  4.16 &  9.5 & 15.1 & 1.4 & 1.2 &  11.15 \\
44(N1530)  &    38.75 & 4500 &  3.93 &  8.5 & 16.5 & 1.2 & 1.1 &  13.25 \\
57(N1530)  &    38.68 & 5000 &  4.47 &  0.1 & 15.6 & 1.0 & 0.9 &  14.92 \\
96(N1530)  &    38.23 & 4700 &  5.41 &  0.2 &  5.0 & 0.4 & 0.3 &  13.20 \\
11(N3359)  &    39.08 & 6300 &  4.16 &  9.5 & 24.2 & 2.7 & 2.3 &   9.04 \\
26(N3359)  &    38.82 & 2200 &  2.22 &  2.7 & 18.7 & 1.5 & 1.3 &  12.88 \\
29(N3359)  &    38.72 & 2800 &  2.76 &  4.2 & 13.5 & 1.2 & 1.0 &  11.55 \\
47(N3359)  &    38.54 & 2600 &  2.94 &  4.7 &  8.9 & 0.8 & 0.7 &  11.62 \\
48(N3359)  &    38.54 & 2000 &  2.40 &  3.2 &  9.3 & 0.8 & 0.7 &  12.09 \\
96(N3359)  &    38.16 & 1600 &  2.50 &  3.4 &  4.6 & 0.3 & 0.3 &  14.36 \\
143(N3359) &    37.93 & 1500 &  2.75 &  4.1 &  4.4 & 0.2 & 0.2 &  23.64 \\
15(N6951)  &    38.74 & 3400 &  3.20 &  5.6 &  8.0 & 1.2 & 1.1 &   6.58 \\
38(N6951)  &    38.36 & 2800 &  3.40 &  6.4 &  7.3 & 0.5 & 0.4 &  14.35 \\
51(N6951)  &    38.30 & 2500 &  3.30 &  6.0 &  4.2 & 0.4 & 0.4 &   9.53 \\
37(N6951)  &    38.37 & 2900 &  3.52 &  6.8 &  5.6 & 0.5 & 0.4 &  10.96 \\
\hline
\end{tabular}
\label{Tabmass}
\end{table*}

\section{Conclusions}

We have analyzed the integrated line profiles of the \hii\ region populations of 
three spiral galaxies NGC~1530, NGC~3359 and NGC~6951 and studied their 
log~L$_{\scriptsize\ha}$--log~$\rm\sigma_{nt}$ relations and have reached the following conclusions:
 
\begin{itemize}

\item The major fraction of the integrated line profiles for the most
  luminous \hii\ regions in the three galaxies show the presence of se\-con\-da\-ry
  Gaussian components. A fraction of $\sim$70\% have two or three Gaussian
  components. These se\-con\-da\-ry components may be blended with the central,
  most intense peak or be located at relatively high velocity with respect to the central
  peak, and in general have low intensity values.  

\item The Luminosity {\it versus} non--thermal velocity dispersion of the
  principal components of the integrated spectra for the most luminous \hii\
  regions in NGC~1530, NGC~3359 and NGC~6951 shows an envelope that can be
  fitted by the li\-near form, ${\rm log~L_{\scriptsize\ha}}=(36.8\pm0.7)+(2.0\pm0.5)~{\rm
    log~\rm\sigma_{nt}}$, where the luminosity of the \hii\ region is taken directly 
  from the calibration ca\-ta\-lo\-gues and ${\rm log~L_{\scriptsize\ha}}=(36.8\pm0.6)+(2.0\pm0.5)~{\rm
    log~\rm\sigma_{nt}}$, where the luminosity is a fraction of the 
  \hii\ region luminosity, determined as the fraction of the line profile integrated
  intensity in the main Gaussian peak of the emission spectrum. 
 
\item We have estimated the stellar mass content within the \hii\ regions and
  found that it represents a small fraction of the total mass of the
  region. Although a precise determination of the fraction is difficult to
  obtain due to the uncertainties involved, an upper limit of 10\% of the
  total mass of the region in stellar form is a conservatively high value that
  takes into account all the uncertainties. 

\item We have compared the virially derived mass with the mass for the \hii\ region
  obtained from the observed \ha\ lu\-mi\-no\-si\-ty for two groups of \hii\ regions in the 
  log~L$_{\scriptsize\ha}$--log~$\rm\sigma_{nt}$ diagram: the \hii\ regions located on the envelope and \hii\
  regions located well above the envelope in velocity dispersion (i.e. well below the envelope in
  luminosity). The vi\-ria\-lly estimated mass for the
  \hii\ regions located well above the envelope are significantly bigger than the masses
  obtained from the \ha\ luminosity, indicating that the velocity dispersions
  for these \hii\ regions include contributions from non equilibrium processes that affect the
  internal ki\-ne\-ma\-tics of these regions. 

\item  For the \hii\ regions located on the envelope, the values for virial masses are
  similar to those for the masses obtained from the \ha\ luminosity when the contamination 
  of other mechanisms in the
  broadening of the central peak is taken into account. This result shows
  that those \hii\ regions in a defined high luminosity range with the lowest
  velocity dispersion can be taken as virialized systems, while the remaining regions are not in virial
  equilibrium.  

\end{itemize}

\appendix
\section{The effects of different  \hii\ region size estimate criteria}

  As noted in Sec.~1 of this article the criteria for estimating the 
size of an \hii\ region vary among authors (see e.g. Rozas et al. 1996; 
McCall et al. 1990, 1996; Sandage \& Tamman 1974) and this prevents valid 
comparisons between different  L--$\sigma$  relations
(e.g. Sandage \& Tamman 1974; Melnick 1977; Gallagher \& Hunter 1983). In order to quantify this 
effect we have compared different sets of quantitative criteria for defining the projected 
area of an \hii\ region and analyzed their effects on the L--$\sigma$ relation.

In the present paper, following a prescription suggested by 
Cepa \& Beckman (1989) and Knapen et al. (1993),  we determine 
the size of the \hii\ regions using the isophote with intensity level
 three times the r.m.s. noise 
level of the local background above the local background intensity level 
(see Sec.~2.2). The integrated flux coming from within this 
isophote is considered  as signal from the \hii\ region, measuring its \ha\ luminosity. To 
make comparison of this method with the alternative based on the use of a fraction 
of the \hii\ region peak intensity to define the limiting isophote,
we have taken  one of the galaxies of the paper, NGC~3359, as a test galaxy. 
We have also varied 
the cut--off level entailed in our own method, thus giving results for a range of different 
possible hypothetical noise le\-vels in the image. The specific criteria for the cut--off isophotes 
are:

\begin{enumerate}
\item Isophote at $1/e$ of the peak intensity of the \hii\ region.
\item Isophote at half peak intensity of the \hii\ region.
\item Isophote at 5 times the  r.m.s. noise  plus the local background level.
\item Isophote at 7 times the  r.m.s. noise  plus the  local background level.
\end{enumerate} 
The luminosity of the \hii\ regions in each case is obtained by integrating the 
emission coming from the pixels inside the co\-rres\-pon\-ding cut--off 
isophote.

\begin{figure}[!h]
\hspace{-0.5cm}
\includegraphics[width=9.2cm]{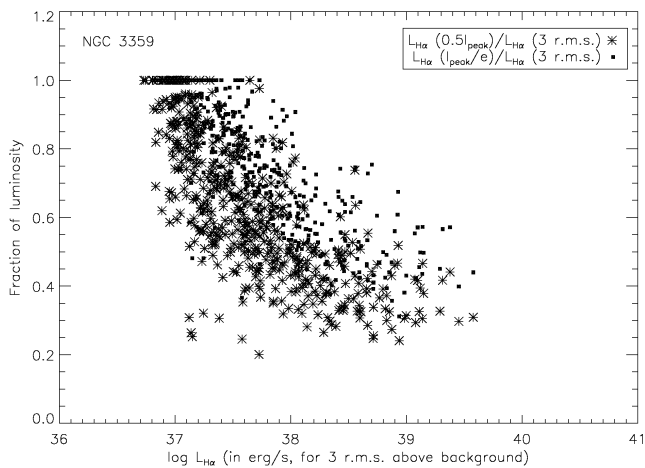}
\protect\caption[ ]{Ratio of the measured luminosity for the \hii\ regions of NGC~3359 using a cut--off isophote 
equal to half (stars) or $1/e$ (filled squares) of the peak intensity to the luminosity measured 
using a cut--off isophote equal to 3 times the  r.m.s. noise  plus the background level of the \ha\ image.}
\label{comparison_l_peak}
\end{figure}

\begin{figure}[!h]
\hspace{-0.5cm}
\includegraphics[width=9.2cm]{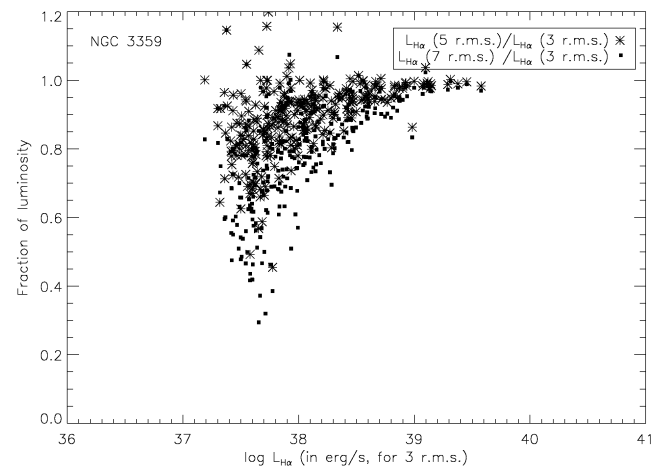}
\protect\caption[ ]{Ratio of the measured luminosity for the \hii\ regions of NGC~3359 using a cut--off isophote 
equal to 5 (stars) and 7 (filled squares) times the r.m.s. noise  plus the background level 
of the \ha\ image of the galaxy, to the luminosity measured using 3 times the r.m.s. noise  plus the background level.}
\label{comparison_sigma}
\end{figure}

The comparison of the \hii\ region luminosity obtained using criteria 
1 and 2 with that obtained using a cut--off at 3 times the r.m.s. noise, adopted 
throughout the paper, is shown in Fig.~\ref{comparison_l_peak} while in Fig.~\ref{comparison_sigma}
we show the comparison of the latter with luminosities obtained using criteria 3 and 4, 
i.e. varying the level above the background for the specified  cut--off.  
We can see in Fig.~\ref{comparison_l_peak} that by integrating to $1/e$ or $1/2$ of 
the peak surface brightness we account for only a fraction of the total \ha\ emission 
detectable above background.
This fraction varies with the luminosity of the region, and the 
shortfall is particularly dramatic for bright regions. Above an \ha\ luminosity of 10$^{38.4}$~erg~s\me\ 
the measured fraction, using this method, lies between 60\% and 20\% of the total detected \ha\ 
above the image noise. The effect is 
readily understood  taking into account that the brighter the \hii\ regions are the 
steeper are the inner sections of their surface brightness profiles (Beckman et al. 2000; Zurita  2001) 
so that a cut--off at a fixed fraction of the peak brightness misses an incresing 
fraction of the total integrated luminosity at higher luminosities. This problem is  progressively 
less serious for regions with lower luminosities, where a cut--off at $1/e$ or $1/2$ of the 
central surface brightness  tends to coincide with a cut--off at 3 times the r.m.s. noise above 
background. 

However the use of a cut--off using this noise--related criterion is not, in 
principle, exempt from problems when com\-pa\-ring results from different authors, as these might come 
from observational data of varying quality and depth. In order to simulate data sets of varying 
quality we have created \ha\ catalogues of \hii\ regions from the same galaxies as 
before, using the same images, but placing our cut--off levels at 5 and 7 times the r.m.s.
instead of at 3 times,  thus exploring the effect which
would accompany varying the signal to noise ratio.

The resulting comparisons are shown in Fig.~\ref{comparison_sigma}
As one would expect, the plot shows considerable dispersion in 
the low luminosity range, decreasing notably towards higher luminosities. For \hii\ regions with 
luminosities greater than 10$^{38.4}$erg~s\me, changing the cut--off criterion from 3 times to 5 
and then 7 times the r.m.s. noise level above the background causes a loss in integrated region 
luminosity of between 2\% and 15\%.
This is directly comparable with losses ranging from 40\% to 80\% 
using the $1/e$ isophote cut--off illustrated in Fig.~\ref{comparison_l_peak}.

Although we believe that the data contained in Figs.~\ref{comparison_l_peak} and \ref{comparison_sigma}
demonstrate that the technique we have adopted is re\-lia\-ble, it is still worth seeing the 
effect of varying the parameters of the different cut--off criteria on the log~L$_{{\rm H}\alpha}$--log~$\sigma_{\rm nt}$ 
relation, using  NGC~3359 as a test galaxy. In Table~\ref{apen_tab_3359} we present the results of this exercise for 
NGC~3359. We can see that varying the cut--off by changing from 3 times to 5 or 7 times the 
r.m.s. noise above background gives no significant change in the slope of the lower envelope to 
the log~L$_{{\rm H}\alpha}$--log~$\sigma_{\rm nt}$ diagram, but there is a slight trend to reduced 
slopes when the alternative criterion using a fraction of the peak brightness as cut--off is used to determine the \ha\ 
flux, which is as expected from Fig.~\ref{comparison_sigma}.

 \begin{table}[!h]
\centering
\caption[]{Slope of the envelope of the log~L$_{{\rm H}\alpha}$-log~$\sigma_{\rm nt}$ relation for the \hii\
regions of NGC~3359 from \hii\ region catalogues obtained using different criteria to estimate the \hii\ region 
cut--off isophote (column 1). 
The slopes (columns 2 and 4) have been obtained with the procedure
described in Sect.~6.2, and each value represents the median value of
the slopes obtained by stepping the starting luminosity in the
luminosity bins by 0.01dex. Columns 2 and 4 show the median value when taking into account only \hii\ 
regions with luminosities greater than  log~L$_{\scriptsize\ha}>$38.4 and 38.8 (in erg~s\me) respectively.
The standard deviation of each set of slopes is given in columns 3 and 5. Column 6 shows the completeness limit 
of the \hii\ region catalogue obtained for each  cut--off isophote criterion.
The symbol ``--'' means that there are not enough points to fit an envelope in that case. }
\begin{tabular}{c|c c|c c|c}
\hline
\hline
Cut--off &  \multicolumn{2}{|c|}{\small for log~L$_{{\rm H}\alpha}$$>38.4$} & \multicolumn{2}{|c|}{\small for log~L$_{{\rm H}\alpha}$$>38.8$ }&Complet.  \\
isophote &  slope & stdev & slope & stdev &limit \\
\hline
I$_{\rm peak}/$e   & 1.18 & 0.08&1.2 &0.2 & 37.4 \\
0.5I$_{\rm peak}$  & 1.2 & 0.2 & -- & --  & 37.4 \\
3 r.m.s.       & 1.4 & 0.3 &1.4 & 0.3 & 37.1 \\
5 r.m.s.       & 1.5 & 0.2 &1.4 & 0.3 & 37.2 \\
7 r.m.s.       & 1.5 & 0.2 &1.4 & 0.3 & 37.4 \\
\hline
\end{tabular}
\label{apen_tab_3359}
\end{table}

We have repeated this exercise for NGC~1530, where 
the image has a considerably lower signal to noise ratio. In this case the slopes obtained using 
all four criteria stated above converge to comparable values. This is because the 
higher noise in the image causes the cut--off levels at 5 and 7 times the r.m.s. noise to approach (and 
even in some cases slightly exceed) the levels at $1/e$ and $1/2$ of the central peak surface 
brightness. We note in Tables~\ref{apen_tab_3359} and \ref{apen_tab_1530} that the completeness limits vary according to the 
cut--off criteria, and that we should be careful not to make comparisons unless the stated lower 
limiting luminosity of our sample is above the completeness limit for the case chosen. Thus 
the slopes cited in column 2 of each table are valid  only for the cut--off levels at I$_{\rm peak}/e$, 
and I$_{\rm peak}/2$,  and at 3 times r.m.s. noise level above the background.

\begin{table}[!h]
\centering
\caption[]{Same as Table~\ref{apen_tab_3359} but for the \hii\ regions of NGC~1530.}
\begin{tabular}{c|c c|c c|c}
\hline
\hline
Cut--off &  \multicolumn{2}{|c|}{\small for log~L$_{{\rm H}\alpha}$$>38.4$} & \multicolumn{2}{|c|}{\small for log~L$_{{\rm H}\alpha}$$>38.8$ }&Complet.  \\
isophote &  slope & stdev & slope & stdev &limit \\
\hline
I$_{\rm peak}/$e  &2.2  &0.2 & 2.4 & 0.3 & 38.2 \\
0.5I$_{\rm peak}$ &2.3  &0.3 & 2.6 & 0.5 & 38.1 \\
3 r.m.s.      &2.0  &0.2 & 2.1 & 0.6 & 38.2 \\
5 r.m.s.      &2.9  &0.4 & 2.5 & 0.4 & 38.4 \\
7 r.m.s.      &3.3  &0.5 & 2.4 & 0.7 & 38.5 \\
\hline
\end{tabular}
\label{apen_tab_1530}
\end{table}
 
We can conclude as a result of the exercises presented in this 
Appendix that the use of different criteria for estimating the areas of \hii\ regions in 
\ha\ images does produce a significant effect on the derived luminosities of the 
regions, an effect which is not constant, nor even linear, but varies systematically with 
the luminosity of the region.
From Figs.~\ref{comparison_l_peak} and \ref{comparison_sigma} we can see that the 
use of a cut--off at three times the r.m.s. noise is the method of choice for defining 
the area of an \hii\ region. It permits more accurate comparisons between authors 
when data of different quality and depth are used. In high sensitivity exposures the 
use of a cut--off based on a fixed fraction of peak intensity leads to  envelopes with 
shallower slopes in a log~L$_{{\rm H}\alpha}$--log~$\sigma_{\rm nt}$ diagram, and this reflects
the fact that this method does not include a significant fraction of the total luminosities
of the most luminous regions. Although the two methods converge 
for images of lower quality this is not su\-ffi\-cient reason to eschew the technique used in 
the present paper.
\\

\begin{acknowledgements}
We thank the anonymous referee for rigorous comments which have
led to significant additions and improvements to the paper.
This work was supported by the Spanish DGES (Direcci\'on General de 
Ense\~nanza Superior) via Grants PB91-0525, PB94-1107 and PB97-0219 
and by the Ministry of Science and Technology via grant AYA2001-0435.  
The WHT and the JKT are ope\-ra\-ted on the island of La Palma by the Isaac Newton Group in
the Spanish Observatorio del Roque de los Muchachos of the Instituto de Astrof\'\i sica de 
Canarias. Partial financial support comes from
the spanish Consejer\'\i a de Educaci\'on y Ciencia de la Junta de Andaluc\'\i a. 
Thanks to Johan Knapen for the narrow band \ha\ image of NGC~1530 and to 
Dan Bramich for his program to subtract the continuum of this image. 
\end{acknowledgements}

\end{document}